%% file: main.tex
\RequirePackage{fix-cm}
\documentclass[twocolumn,epjc3,runningheads,smartrunhead]{svjour3}  
\pdfoutput=1
\smartqed  
\RequirePackage[colorlinks,breaklinks]{hyperref}
\hypersetup{linkcolor=blue,citecolor=blue,filecolor=black,urlcolor=blue} 
\usepackage{graphicx}
\usepackage{atlasphysics}
\usepackage{mathrsfs}
\usepackage{amssymb}
\usepackage{subfigure}
\usepackage{xspace}
\usepackage{preprintcover} 
\usepackage{cite,mcite} 
\journalname{Eur.\ Phys.\ J.\ C}
\PreprintCoverPaperTitle{%
  Measurement of the Top Quark Mass with the Template Method
  in the $\mathbf{t\bar{t} \to} \mbox{lepton+jets}$ Channel using
  ATLAS Data}
\PreprintIdNumber{CERN-PH-EP-2012-003}
\PreprintCoverAbstract{%
 The top quark mass has been measured using the template method in the
 \ttbarlj\ channel based on data recorded in 2011 with the ATLAS detector at the
 LHC.
 The data were taken at a proton-proton centre-of-mass energy of
 $\rts=7$~\TeV\ and correspond to an integrated luminosity of \atlumo~\ifb.
 The analyses in the \ejets\ and \mjets\ decay channels yield consistent
 results.
 The top quark mass is measured to be $\mt=\mttd$~\GeV.
}
\PreprintJournalName{Eur.\ Phys.\ J.\ C} 
\DeclareGraphicsExtensions{.pdf}
\newcommand{\atlumo    }{\ensuremath{1.04}\xspace}
\newcommand{\atlumounc }{\ensuremath{3.7\%}\xspace}
\newcommand{\sigttbar  }{\ensuremath{164.6~\mathrm{pb}}\xspace}
\newcommand{\dsigttbar }{\ensuremath{8\%}\xspace}
\newcommand{\mttevo  }{\ensuremath{\XZ{173.2}{0.6}{0.8}}\xspace}
\newcommand{\tevlum  }{\ensuremath{5.8}\xspace}
\newcommand{\mtatoodelval }{\ensuremath{172.93}\xspace}
\newcommand{\mtatoodelsta }{\ensuremath{1.46}\xspace}
\newcommand{\mtatoodelsys }{\ensuremath{2.46}\xspace}
\newcommand{\mtatoodel    }{\ensuremath{\XZ{172.9}{1.5}{2.5}}\xspace}
\newcommand{\mtatoodmuval }{\ensuremath{175.54}\xspace}
\newcommand{\mtatoodmusta }{\ensuremath{1.13}\xspace}
\newcommand{\mtatoodmusys }{\ensuremath{2.56}\xspace}
\newcommand{\mtatoodmu    }{\ensuremath{\XZ{175.5}{1.1}{2.6}}\xspace}
\newcommand{\mtatotdelval }{\ensuremath{174.30}\xspace}
\newcommand{\mtatotdelsta }{\ensuremath{0.83}\xspace}
\newcommand{\mtatotdelsys }{\ensuremath{2.31}\xspace}
\newcommand{\mtatotdel    }{\ensuremath{\XZ{174.3}{0.8}{2.3}}\xspace}
\newcommand{\mtatotdelori }{\ensuremath{\XZtd{174.3}{1.0}{2.2}}\xspace}
\newcommand{\mtatotdmuval }{\ensuremath{175.01}\xspace}
\newcommand{\mtatotdmusta }{\ensuremath{0.74}\xspace}
\newcommand{\mtatotdmusys }{\ensuremath{2.57}\xspace}
\newcommand{\mtatotdmu    }{\ensuremath{\XZ{175.0}{0.7}{2.6}}\xspace}
\newcommand{\mtatotdmuori }{\ensuremath{\XZtd{175.0}{0.9}{2.5}}\xspace}
\newcommand{\jesatotdelval}{\ensuremath{0.985}\xspace}
\newcommand{\jesatotdelsta}{\ensuremath{0.008}\xspace}
\newcommand{\jesatotdmdval}{\ensuremath{0.986}\xspace}
\newcommand{\jesatotdmdsta}{\ensuremath{0.006}\xspace}
\newcommand{\cortd        }{\ensuremath{-0.57}\xspace}
\newcommand{\mtatostacut }{\ensuremath{0.6}\xspace}
\newcommand{\mtatosyscut }{\ensuremath{2.3}\xspace}
\newcommand{\statcorel   }{\ensuremath{0.15}\xspace}
\newcommand{\statcormu   }{\ensuremath{0.16}\xspace}
\newcommand{\mtodval   }{\ensuremath{174.35}\xspace}
\newcommand{\mtodsta   }{\ensuremath{0.91}\xspace}
\newcommand{\mtodsys   }{\ensuremath{2.50}\xspace}
\newcommand{\mtod      }{\ensuremath{\XZ{174.4}{0.9}{2.5}}\xspace}
\newcommand{\mttdval   }{\ensuremath{174.53}\xspace}
\newcommand{\mttdsta   }{\ensuremath{0.61}\xspace}
\newcommand{\mttdsys   }{\ensuremath{2.31}\xspace}
\newcommand{\mttd      }{\ensuremath{\XZ{174.5}{0.6}{2.3}}\xspace}
\newcommand{\rhomin    }{\ensuremath{0.63}\xspace}
\newcommand{\rhomax    }{\ensuremath{0.77}\xspace}
\newcommand{\difftot }{\ensuremath{16\%}\xspace}
\newcommand{\fraco   }{\ensuremath{50\%}\xspace}
\newcommand{\maxsys  }{\ensuremath{85\%}\xspace}
\newcommand{\isoenel      }{\ensuremath{3.5}\xspace}
\newcommand{\isoenmu      }{\ensuremath{4}\xspace}
\newcommand{\btageff      }{\ensuremath{70\%}\xspace}
\newcommand{\btagrej      }{\ensuremath{100}\xspace}
\newcommand{\Nevelod      }{\ensuremath{1151}\xspace}
\newcommand{\Nevmuod      }{\ensuremath{1724}\xspace}
\newcommand{\Neveltd      }{\ensuremath{4556}\xspace}
\newcommand{\Nevmutd      }{\ensuremath{7225}\xspace}
\newcommand{\nprimver     }{\ensuremath{\mbox{four}}\xspace}
\newcommand{\Wbtag        }{\ensuremath{W_{\mathrm{btag}}}\xspace}
\newcommand{\fullmalow    }{\ensuremath{30\%}\xspace}
\newcommand{\fullmahig    }{\ensuremath{40\%}\xspace}
\newcommand{\jetcorper    }{\ensuremath{70\%}\xspace}
\newcommand{\mtstaelod    }{\ensuremath{1.36 \pm 0.16}\xspace}
\newcommand{\mtstamuod    }{\ensuremath{1.11 \pm 0.06}\xspace}
\newcommand{\buncpro      }{\ensuremath{50\%}\xspace}
\newcommand{\lnL          }{\ensuremath{\ln\,L}\xspace}
\newcommand{\likecut      }{\ensuremath{-50}\xspace}
\newcommand{\Ejet         }[1]{\ensuremath{E_{\mathrm{jet}_{#1}}}\xspace}
\newcommand{\tf           }{\ensuremath{{\mathcal{T}}}\xspace}
\newcommand{\bw           }{\ensuremath{{\mathcal{B}}}\xspace}
\newcommand{\Emis         }[1]{\ensuremath{E^{\mathrm{miss}}_{#1}}\xspace}
\newcommand{\chielod      }{\ensuremath{21}\xspace}
\newcommand{\ndfelod      }{\ensuremath{23}\xspace}
\newcommand{\chimuod      }{\ensuremath{39}\xspace}
\newcommand{\ndfmuod      }{\ensuremath{23}\xspace}
\newcommand{\mtstaeltd     }{\ensuremath{1.20 \pm 0.08}\xspace}
\newcommand{\mtstamutd     }{\ensuremath{0.94 \pm 0.04}\xspace}
\newcommand{\JSF           }{\ensuremath{\mbox{JSF}}\xspace}
\newcommand{\JSFmax        }{\ensuremath{\pm 2.5\%}\xspace}
\newcommand{\mwmtopcortd   }{\ensuremath{-0.06}\xspace}
\newcommand{\chieltd       }{\ensuremath{47}\xspace}
\newcommand{\ndfeltd       }{\ensuremath{38}\xspace}
\newcommand{\chimutd       }{\ensuremath{51}\xspace}
\newcommand{\ndfmutd       }{\ensuremath{38}\xspace}
\newcommand{\mtstasysmin   }{\ensuremath{0.2}\xspace}
\newcommand{\mtstasysmax   }{\ensuremath{0.5}\xspace}
\newcommand{\sysWnorm      }{\ensuremath{\pm 70\%}\xspace}
\newcommand{\uncqnorm      }{\ensuremath{\pm 100\%}\xspace}
\newcommand{\uncjesglobmin }{\ensuremath{\pm 2.5\%}\xspace}
\newcommand{\uncjesglobmax }{\ensuremath{\pm 8\%}\xspace}
\newcommand{\uncpilecent   }{\ensuremath{\pm 2.5\%}\xspace}
\newcommand{\uncpileforw   }{\ensuremath{\pm 5\%}\xspace}
\newcommand{\uncbjesmin    }{\ensuremath{\pm 0.8\%}\xspace}
\newcommand{\uncbjesmax    }{\ensuremath{\pm 2.5\%}\xspace}
\newcommand{\uncbrel       }{\ensuremath{\pm 2\%}\xspace}
\newcommand{\uncjeff       }{\ensuremath{\pm 2\%}\xspace}
\newcommand{\od    }{\ensuremath{\mbox{1d-analysis}}\xspace}
\newcommand{\td    }{\ensuremath{\mbox{2d-analysis}}\xspace}
\newcommand{\rof   }{\ensuremath{\rho}\xspace}
\newcommand{\X            }[2]{\ensuremath{#1\,\pm #2}\xspace}
\newcommand{\XZ           }[3]{\ensuremath{#1\,\pm #2_\mathrm{stat}\,\pm #3_\mathrm{syst}}\xspace}
\newcommand{\XZtd         }[3]{\ensuremath{#1\,\pm #2_\mathrm{stat+JSF}\,\pm #3_\mathrm{syst}}\xspace}
\newcommand{\ttbarlj      }{\ensuremath{\ttbar\to\mbox{lepton+jets}}\xspace}
\newcommand{\ejets        }{\ensuremath{e\mbox{+jets}}\xspace}
\newcommand{\mjets        }{\ensuremath{\mu\mbox{+jets}}\xspace}
\newcommand{\blep         }{\ensuremath{b_{\ell}}\xspace}
\newcommand{\bhad         }{\ensuremath{b_\mathrm{had}}\xspace}
\newcommand{\qone         }{\ensuremath{q_1}\xspace}
\newcommand{\qtwo         }{\ensuremath{q_2}\xspace}
\newcommand{\Wj           }{\ensuremath{\Wboson\mbox{+jets}}\xspace}
\newcommand{\Zj           }{\ensuremath{\Zboson\mbox{+jets}}\xspace}

\newcommand{\Gt           }{\ensuremath{\Gamma_{\mathrm{top}}}\xspace}
\newcommand{\mt           }{\ensuremath{m_{\mathrm{top}}}\xspace}
\newcommand{\mtr          }{\ensuremath{\mt^{\mathrm{reco}}}\xspace}
\newcommand{\mtrl         }{\ensuremath{\mt^{\mathrm{reco,like}}}\xspace}
\newcommand{\RtW          }{\ensuremath{R_{\mathrm{32}}}\xspace}

\newcommand{\mWt          }{\ensuremath{m^{\mathrm{T}}_{\mathrm{\Wboson}}}\xspace}
\newcommand{\mWr          }{\ensuremath{m_{\mathrm{W}}^{\mathrm{reco}}}\xspace}
\newcommand{\mWrl         }{\ensuremath{m_{\mathrm{W}}^{\mathrm{reco,like}}}\xspace}
\newcommand{\etaclus      }{\ensuremath{\eta_\mathrm{cluster}}\xspace}
\newcommand{\absetaclus   }{\ensuremath{\vert\etaclus\vert}\xspace}
\newcommand{\Psig         }{\ensuremath{P_{\mathrm{sig}}}\xspace}
\newcommand{\Pbkg         }{\ensuremath{P_{\mathrm{bkg}}}\xspace}
\newcommand{\Ptop         }{\ensuremath{P_{\mathrm{top}}}\xspace}
\newcommand{\Ptopsig      }{\ensuremath{P_{\mathrm{top}}^{\mathrm{sig}}}\xspace}
\newcommand{\Ptopbkg      }{\ensuremath{P_{\mathrm{top}}^{\mathrm{bkg}}}\xspace}
\newcommand{\PW           }{\ensuremath{P_{\mathrm{W}}}\xspace}
\newcommand{\PWsig        }{\ensuremath{P_{\mathrm{W}}^{\mathrm{sig}}}\xspace}
\newcommand{\PWbkg        }{\ensuremath{P_{\mathrm{W}}^{\mathrm{bkg}}}\xspace}
\newcommand{\chiq         }{\ensuremath{\chi^2}\xspace}
\newcommand{\chidof       }{\ensuremath{\chi^2/\mathrm{dof}}\xspace}
\newcommand{\ali          }[1]{\ensuremath{\alpha_{\mathrm{#1}}}\xspace}
\newcommand{\nsig         }{\ensuremath{n_{\mathrm{sig}}}\xspace}
\newcommand{\nbkg         }{\ensuremath{n_{\mathrm{bkg}}}\xspace}
\newcommand{\lami         }{\ensuremath{\lambda_{\mathrm{i}}}\xspace}
\newcommand{\Ni           }{\ensuremath{N_{\mathrm{i}}}\xspace}
\newcommand{\Nbins        }{\ensuremath{N_{\mathrm{bins}}}\xspace}
\newcommand{\nbpred       }{\ensuremath{\nbkg^{\mathrm{pred}}}\xspace}

\newcommand{\Like         }[1]{\ensuremath{\mathcal{L}_{\mathrm{#1}}}\xspace}

\newcommand{\btag         }{\ensuremath{b\mbox{-tagging}}\xspace}
\newcommand{\bjet         }{\ensuremath{b\mbox{-jet}}\xspace}
\newcommand{\bjets        }{\ensuremath{b\mbox{-jets}}\xspace}
\newcommand{\bquark       }{\ensuremath{b\mbox{-quark}}\xspace}
\newcommand{\bquarks      }{\ensuremath{b\mbox{-quarks}}\xspace}
\newcommand{\bhadrons     }{\ensuremath{b\mbox{-hadrons}}\xspace}
\newcommand{\kt           }{\ensuremath{k_\mathrm{t}}\xspace}
\newcommand{\Acermc       }{{\sc AcerMC}\xspace}
\newcommand{\Alpgen       }{{\sc Alpgen}\xspace}
\newcommand{\Herwig       }{{\sc Herwig}\xspace}
\newcommand{\Jimmy        }{{\sc Jimmy}\xspace}
\newcommand{\Mcatnlo      }{{\sc MC@NLO}\xspace}
\newcommand{\Pythia       }{{\sc Pythia}\xspace}
\newcommand{\Powheg       }{{\sc Powheg}\xspace}
\begin{document}
\title{Measurement of the Top Quark Mass with the Template Method in the
  $\mathbf{t\bar{t}} \to \mbox{lepton+jets}$ Channel using ATLAS Data} 

\titlerunning{Measurement of the Top Quark Mass with the Template Method
  using ATLAS Data}

\author{The ATLAS Collaboration}

\institute{CERN, 1211 Geneva 23, Switzerland,\\
  \email{atlas.publications@cern.ch}}

\date{Received: date / Accepted: date}
\maketitle
%
%
\begin{abstract}
 The top quark mass has been measured using the template method in the
 \ttbarlj\ channel based on data recorded in 2011 with the ATLAS detector at the
 LHC.
 The data were taken at a proton-proton centre-of-mass energy of
 $\rts=7$~\TeV\ and correspond to an integrated luminosity of \atlumo~\ifb.
 The analyses in the \ejets\ and \mjets\ decay channels yield consistent
 results.
 The top quark mass is measured to be $\mt=\mttd$~\GeV.
 \keywords{Top quark mass \and Template method \and ATLAS}
\end{abstract}
%
\section{Introduction}
\label{sec:intro}
%
 The top quark mass (\mt) is a fundamental parameter of the Standard Model (SM)
 of particle physics.
 Due to its large mass, the top quark gives large contributions to electroweak
 radiative corrections. Together with precision electroweak measurements, the
 top quark mass can be used to derive constraints on the masses of the as yet
 unobserved Higgs boson~\cite{TEV-1102,CON-2011-157}, and of heavy particles
 predicted by extensions of the SM.
 After the discovery of the top quark in 1995, much work has been devoted to the
 precise measurement of its mass. The present average value of
 $\mt=\mttevo$~GeV~\cite{TEV-1101} is obtained from measurements at the Tevatron
 performed by CDF and D$\emptyset$ with Run I and Run II data corresponding to
 integrated luminosities of up to \tevlum~\ifb.
 At the LHC, \mt\ has been measured by CMS in \ttbar\ events in which both
 \Wboson\ bosons from the top quark decays themselves decay into a charged
 lepton and a neutrino~\cite{CMS-1101}.

 \begin{sloppypar}
 The main methodology used to determine \mt\ at hadron colliders consists of
 measuring the invariant mass of the decay products of the top quark candidates
 and deducing \mt\ using sophisticated analysis methods.
 The most precise measurements of this type use the \ttbarlj\ channel, i.e.~the
 decay $\ttbar \to \ell\nu\blep\,\qone\qtwo\bhad$ with $\ell=e,\mu$, where one
 of the \Wboson\ bosons from the \ttbar\ decay decays into a charged lepton and
 a neutrino and the other into a pair of quarks, and where \blep(\bhad) denotes
 the \bquark\ associated to the leptonic (hadronic) \Wboson\ boson decay.
 In this paper these \ttbar\ decay channels are referred to as \ejets\ and
 \mjets\ channels.

 In the template method, simulated distributions are constructed for a chosen
 quantity sensitive to the physics observable under study, using a number of
 discrete values of that observable.
 These templates are fitted to functions that interpolate between different
 input values of the physics observable, fixing all other parameters of the
 functions.
 In the final step a likelihood fit to the observed data distribution is used to
 obtain the value for the physics observable that best describes the data.
 In this procedure, the experimental distributions are constructed such that
 they are unbiased estimators of the physics observable used as an input
 parameter in the signal Monte Carlo samples.
 Consequently, the top quark mass determined this way from data corresponds
 to the mass definition used in the Monte Carlo.
 It is expected~\cite{BUC-1101} that the difference between this mass definition
 and the pole mass is of order 1~GeV.
 \end{sloppypar}

 The precision of the measurement of \mt\ is limited mainly by the systematic
 uncertainty from a few sources.
 In this paper two different estimators for \mt\ are developed, which have only
 a small statistical correlation and use different strategies to reduce the
 impact of these sources on the final uncertainty. This choice translates into
 different sensitivities to the uncertainty sources for the two estimators.
 The first implementation of the template method is a one-dimensional template
 analysis (\od), which is based on the observable \RtW, defined as the per event
 ratio of the reconstructed invariant masses of the top quark and the
 \Wboson\ boson reconstructed from three and two jets respectively.
 For each event, an event likelihood is used to select the jet triplet assigned
 to the hadronic decays of the top quark and the \Wboson\ boson amongst the jets
 present in the event.
 The second implementation is a two-dimensional template analysis (\td), which
 simultaneously determines \mt\ and a global jet energy scale factor (JSF) from
 the reconstructed invariant masses of the top quark and the \Wboson\ boson.
 This method utilises a \chiq\ fit that constrains the reconstructed invariant
 mass of the \Wboson\ boson candidate to the world-average
 \Wboson\ boson mass measurement~\cite{NAK-1001}.

 The paper is organised as follows: details of the ATLAS detector are given in
 Section~\ref{sec:atlas}, the data and Monte Carlo simulation samples are
 described in Section~\ref{sec:mc}. The common part of the event selections is
 given in Section~\ref{sec:evsel}, followed by analysis-specific requirements
 detailed in Section~\ref{sec:addreq}.
 The specific details of the two analyses are explained in
 Section~\ref{sec:1dim} and Section~\ref{sec:2dim}.
 The measurement of \mt\ is given in Section~\ref{sec:measure}, where the
 evaluation of the systematic uncertainties is discussed in
 Section~\ref{sec:syserr}, and the individual results and their combination are
 reported in Section~\ref{sec:rescom}.
 Finally, the summary and conclusions are given in Section~\ref{sec:summar}.
%
\section{The ATLAS detector}
\label{sec:atlas}
%
 The ATLAS detector~\cite{ATL-2008-001} at the LHC covers nearly the entire
 solid angle around the collision point\footnote{ATLAS uses a right-handed
   coordinate system with its origin at the nominal interaction point (IP) in
   the centre of the detector and the $z$-axis along the beam pipe. The $x$-axis
   points from the IP to the centre of the LHC ring, and the $y$ axis points
   upward. Cylindrical coordinates $(r,\phi)$ are used in the transverse plane,
   $\phi$ being the azimuthal angle around the beam pipe. The pseudorapidity is
   defined in terms of the polar angle $\theta$ as
   $\eta=-\ln\tan(\theta/2)$. Transverse momentum and energy are defined as $\pt
   = p\sin\theta$ and $\ET = E\sin\theta$, respectively.}.
 It consists of an inner tracking detector surrounded by a thin superconducting
 solenoid, electromagnetic and hadronic calorimeters, and an external muon
 spectrometer incorporating three large superconducting toroid magnet
 assemblies.

 The inner-detector system is immersed in a 2T axial magnetic field and provides
 charged particle tracking in the range $\abseta < 2.5$. The high-granularity
 silicon pixel detector covers the vertex region and provides typically three
 measurements per track, followed by the silicon microstrip tracker which
 provides four measurements from eight strip layers. These silicon detectors are
 complemented by the transition radiation tracker, which enables extended track
 reconstruction up to $\abseta = 2.0$. In giving typically more than 30
 straw-tube measurements per track, the transition radiation tracker improves
 the inner detector momentum resolution, and also provides electron
 identification information.

 The calorimeter system covers the pseudorapidity range $\abseta < 4.9$. Within
 the region $\abseta < 3.2$, electromagnetic calorimetry is provided by barrel
 and end cap lead/liquid argon (LAr) electromagnetic calorimeters, with an
 additional thin LAr presampler covering $\abseta < 1.8$ to correct for energy
 loss in material upstream of the calorimeters. Hadronic calorimetry is provided
 by the steel/scintillating-tile calorimeter, segmented into three barrel
 structures within $\abseta < 1.7$, and two copper/LAr hadronic endcap
 calorimeters. The solid angle coverage is completed with forward copper/LAr and
 tungsten/LAr calorimeter modules optimised for electromagnetic and hadronic
 measurements respectively.

 The muon spectrometer comprises separate trigger and high-precision tracking
 chambers measuring the deflection of muons in a magnetic field with a bending
 integral up to 8~Tm in the central region, generated by three superconducting
 air-core toroids. The precision chamber system covers the region $\abseta <
 2.7$ with three layers of monitored drift tubes, complemented by cathode strip
 chambers in the forward region. The muon trigger system covers the range
 $\abseta < 2.4$ with resistive plate chambers in the barrel, and thin gap
 chambers in the endcap regions.

 A three-level trigger system is used. The first level trigger is implemented in
 hardware and uses a subset of detector information to reduce the event rate to
 a design value of at most 75~kHz. This is followed by two software-based
 trigger levels, which together reduce the event rate to about 300~Hz.
%
\section{Data and Monte Carlo samples}
\label{sec:mc}
%
 In this paper, data from LHC proton-proton collisions are used, collected at a
 centre-of-mass energy of $\rts=7$~\TeV\ with the ATLAS detector during
 March-June 2011. An integrated luminosity of \atlumo~\ifb\ is included.

 \begin{sloppypar}
 Simulated \ttbar\ events and single top quark production are both generated
 using the Next-to-Leading Order (NLO) Monte Carlo program
 \Mcatnlo~\cite{FRI-0201,FRI-0301} with the NLO parton density function set
 CTEQ6.6~\cite{NAD-0801}.
 Parton showering and underlying event (i.e.~additional interactions of the
 partons within the protons that underwent the hard interaction) are modelled
 using the \Herwig~\cite{COR-0001} and \Jimmy~\cite{BUT-9601} programs.
 For the construction of signal templates, the \ttbar\ and single top quark
 production samples are generated for different assumptions on \mt\ using
 six values (in~\GeV) namely~$(160, 170, 172.5, 175, 180, 190)$, and with the
 largest samples at $\mt=172.5$~\GeV.
 All \ttbar\ samples are normalised to the corresponding cross-sections,
 obtained with the latest theoretical computation approximating the NNLO
 prediction and implemented in the HATHOR package~\cite{ALI-1101}.
 The predicted \ttbar\ cross-section for a top quark mass of
 $\mt=172.5$~\GeV\ is \sigttbar, with an uncertainty of about \dsigttbar.
 \end{sloppypar}
%
\begin{table*}[tbp!]
\begin{center}
\begin{tabular}{|l|r@{$\,\pm\,$}r|r@{$\,\pm\,$}r|r@{$\,\pm\,$}r|r@{$\,\pm\,$}r|} \hline
         & \multicolumn{4}{|c|}{\od} 
         & \multicolumn{4}{|c|}{\td} \\\cline{2-9} 
 Process & \multicolumn{2}{|c|}{\ejets} & \multicolumn{2}{|c|}{\mjets} 
         & \multicolumn{2}{|c|}{\ejets} & \multicolumn{2}{|c|}{\mjets} \\\hline
 \ttbar\ signal      &  990 &   40 &  1450 &   50 & 3400 & 200 & 5100 & 300 \\\cline{6-9}
 Single top (signal) &   43 &    2 &    53 &    3 &  190 &  10 &  280 &  20 \\\cline{2-5}
 \Zj\                &   12 &    3 &     8 &    3 &   83 &   8 &  100 &   8 \\ 
$\Zboson\Zboson /\Wboson\Zboson /\Wboson\Wboson$ 
		     &    2 & $<$1 &     2 & $<$1 &   11 &   2 &   18 &   2 \\
 \Wj\ (data)         &   80 &   60 &   100 &   70 &  700 & 500 & 1100 & 800 \\
 QCD multijet (data) &   50 &   50 &    40 &   40 &  200 & 200 &  400 & 400 \\\hline
 Signal + background & 1180 &   80 &  1650 &   80 & 4500 & 500 & 6900 & 900 \\
 \hline
 Data                & \multicolumn{2}{|c|}{\Nevelod} & \multicolumn{2}{|c|}{\Nevmuod} 
                     & \multicolumn{2}{|c|}{\Neveltd} & \multicolumn{2}{|c|}{\Nevmutd}
                                                                             \\\hline
\end{tabular}
\end{center}
\caption{The observed numbers of events in the data in the \ejets\ and
  \mjets\ channels, for the two analyses after the common event selection and
  additional analysis-specific requirements. In addition, the expected numbers
  of signal and background events corresponding to the integrated luminosity of
  the data are given, where the single top quark production events are treated
  as signal for the \od, and as background for the \td.
  The Monte Carlo estimates assume SM cross-sections. The \Wj\ and QCD multijet
  background contributions are estimated from ATLAS data. The uncertainties for
  the estimates include different components detailed in the text. All predicted
  event numbers are quoted using one significant digit for the uncertainties,
  i.e.~the trailing zeros are insignificant.
\label{tab:cutflow}}
\end{table*}

 \begin{sloppypar}
 The production of \Wboson\ bosons or \Zboson\ bosons in association with jets
 is simulated using the \Alpgen\ generator~\cite{MAN-0301} interfaced to the
 \Herwig\ and \Jimmy\ packages.
 Diboson production processes (\Wboson\Wboson, \Wboson\Zboson\ and
 \Zboson\Zboson) are produced using the \Herwig\ generator.
 All Monte Carlo samples are generated with additional multiple soft
 proton-proton interactions. These simulated events are re-weighted such that
 the distribution of the number of interactions per bunch crossing (pileup) in
 the simulated samples matches that in the data.
 The mean number of primary vertices per bunch crossing for the data of this
 analysis is about \nprimver.
 The samples are then processed through the GEANT4~\cite{AGO-0301}
 simulation~\cite{ATL-2010-005} and the reconstruction software of the ATLAS
 detector.
 \end{sloppypar}
%
\section{Event selection}
\label{sec:evsel}
%
 In the signal events the main reconstructed objects in the detector are
 electron and muon candidates as well as jets and missing transverse momentum
 (\met).
 An electron candidate is defined as an energy deposit in the electromagnetic
 calorimeter with an associated well-reconstructed track.
 Electron candidates are required to have transverse energy $\ET>25$~\GeV\ and
 $\absetaclus < 2.47$, where \etaclus\ is the pseudorapidity of the
 electromagnetic cluster associated with the electron.
 Candidates in the transition region between the barrel and end-cap calorimeter,
 i.e.~candidates fulfilling $1.37<\absetaclus<1.52$, are excluded.
 Muon candidates are reconstructed from track segments in different layers of
 the muon chambers. These segments are combined starting from the outermost
 layer, with a procedure that takes material effects into account, and matched
 with tracks found in the inner detector. The final candidates are refitted
 using the complete track information, and are required to satisfy
 $\pt>20$~\GeV\ and $\vert\eta\vert<2.5$.
 Isolation criteria, which restrict the amount of energy deposits near the
 candidates, are applied to both electron and muon candidates to reduce the
 background from hadrons mimicking lepton signatures and backgrounds from heavy
 flavour decays inside jets. 
 For electrons, the energy not associated to the electron cluster and contained
 in a cone of $\Delta R = \sqrt{{\Delta \phi}^2+{\Delta \eta}^2}=0.2$ must not
 exceed \isoenel~\GeV, after correcting for energy deposits from pileup, which
 in the order of 0.5~\GeV.
 For muons, the sum of track transverse momenta and the total energy deposited
 in a cone of $\Delta R=0.3$ around the muon are both required to be less than
 \isoenmu~\GeV.

 Jets are reconstructed with the anti-\kt algorithm~\cite{CAC-0801} with
 $R=0.4$, starting from energy clusters of adjacent calorimeter cells called
 topological clusters~\cite{LARG-PUB-2008-002}.
 These jets are calibrated first by correcting the jet energy using the scale
 established for electromagnetic objects (EM scale) and then performing a
 further correction to the hadronic energy scale using correction factors, that
 depend on energy and $\eta$, obtained from simulation and validated with
 data~\cite{ATL-2011-085}.
 Jet quality criteria~\cite{CON-2012-020} are applied to identify and reject
 jets reconstructed from energies not associated to energy deposits in the
 calorimeters originating from particles emerging from the bunch crossing under
 study.
 The jets failing the quality criteria, which may have been reconstructed from
 various sources such as calorimeter noise, non-collision beam-related
 background, and cosmic-ray induced showers, can efficiently be
 identified~\cite{CON-2012-020}.

 The reconstruction of \met\ is based upon the vector sum of calorimeter energy
 deposits projected onto the transverse plane.
 It is reconstructed from topological clusters, calibrated at the EM scale and
 corrected according to the energy scale of the associated physics
 object. Contributions from muons are included by using their momentum measured
 from the track and muon spectrometer systems in the \met\ reconstruction.

 Muons reconstructed within a $\Delta R=0.4$ cone of a jet satisfying
 $\pt>20$~\GeV\ are removed to reduce the contamination caused by muons from
 hadron decays within jets. Subsequently, jets within $\Delta R=0.2$ of an
 electron candidate are removed to avoid double counting, which can occur
 because electron clusters are usually also reconstructed as jets.

 Reconstruction of top quark pair events is facilitated by the ability to tag
 jets originating from the hadronisation of \bquarks.
 For this purpose, a neural-net--based algorithm~\cite{CON-2011-102}, relying on
 vertex properties such as the decay length significance, is applied. The chosen
 working point of the algorithm corresponds to a \btag\ efficiency of
 \btageff\ for jets originating from \bquarks\ in simulated \ttbar\ events and a
 light quark jet rejection factor of about \btagrej.
 Irrespective of their origin, jets tagged by this algorithm are called
 \bjets\ in the following, whereas those not tagged are called light jets.

 The signal is characterised by an isolated lepton with relatively high \pt,
 \met\ arising from the neutrino from the leptonic \Wboson\ boson decay, two
 \bquark\ jets, and two light quark jets from the hadronic \Wboson\ boson decay.
 The selection of events consists of a series of requirements on general event
 quality and the reconstructed objects designed to select the event topology
 described above.
 The following event selections are applied:
%
\begin{itemize}
 \item it is required that the appropriate single electron or single muon
   trigger has fired (with thresholds at 20~\GeV\ and 18~\GeV, respectively);
 \item the event must contain one and only one reconstructed lepton with $\ET>
   25$~\GeV\ for electrons and $\pt > 20$~\GeV\ for muons which, for the
   \ejets\ channel, should also match the corresponding trigger object;
 \item in the \mjets\ channel, $\met>20$~\GeV\ and in addition
   $\met+\mWt>60$~\GeV\ is required\footnote{Here \mWt\ is the $W$-boson
   transverse mass, defined as
   $\sqrt{2\,p_\mathrm{T,\ell}\,p_\mathrm{T,\nu}\left[1-\cos(\phi_{\ell}-\phi_{\nu})\right]}$,
   where the measured \met\ vector provides the neutrino ($\nu$) information.};
 \item in the \ejets\ channel more stringent cuts on \met\ and \mWt\ are
   required because of the higher level of QCD multijet background, these being
   $\met > 35$~\GeV\ and $\mWt>25$~\GeV;
 \item the event is required to have $\ge 4$ jets with $\pt>25$~\GeV\ and
   $\vert\eta\vert<2.5$. It is required that at least one of these jets is a
   \bjet.
\end{itemize}
%
 \begin{sloppypar}
 This common event selection is augmented by additional analysis-specific
 event requirements described next.
 \end{sloppypar}
%
\section{Specific event requirements}
\label{sec:addreq}
%
 To optimise the expected total uncertainty on \mt, some specific requirements
 are used in addition to the common event selection.
%
\begin{figure*}[tbp!]
\centering
\subfigure[\ejets\ channel]{
  \includegraphics[width=0.38\textwidth]{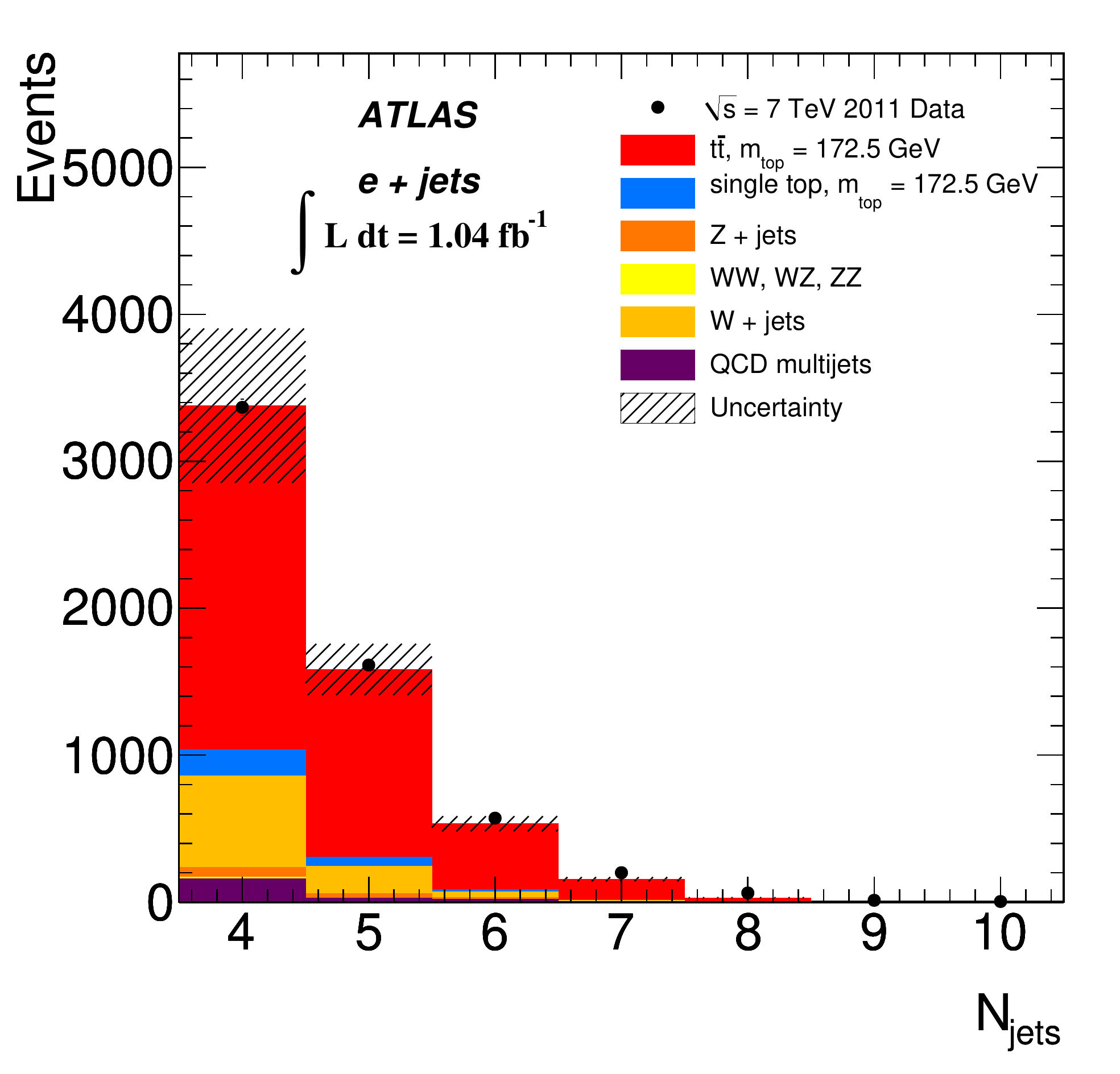}}
\subfigure[\mjets\ channel]{
  \includegraphics[width=0.38\textwidth]{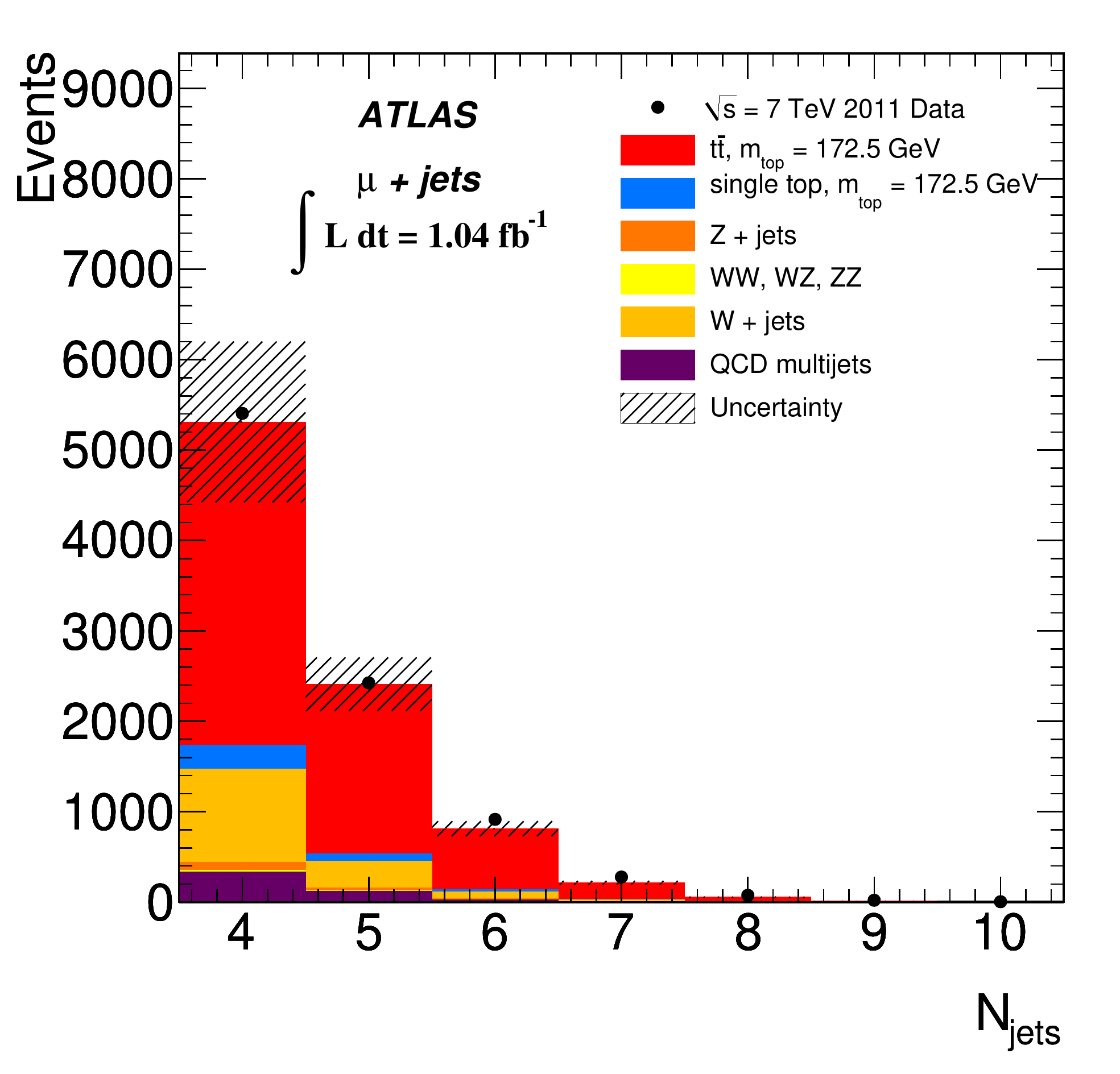}}
\subfigure[\ejets\ channel]{
  \includegraphics[width=0.38\textwidth]{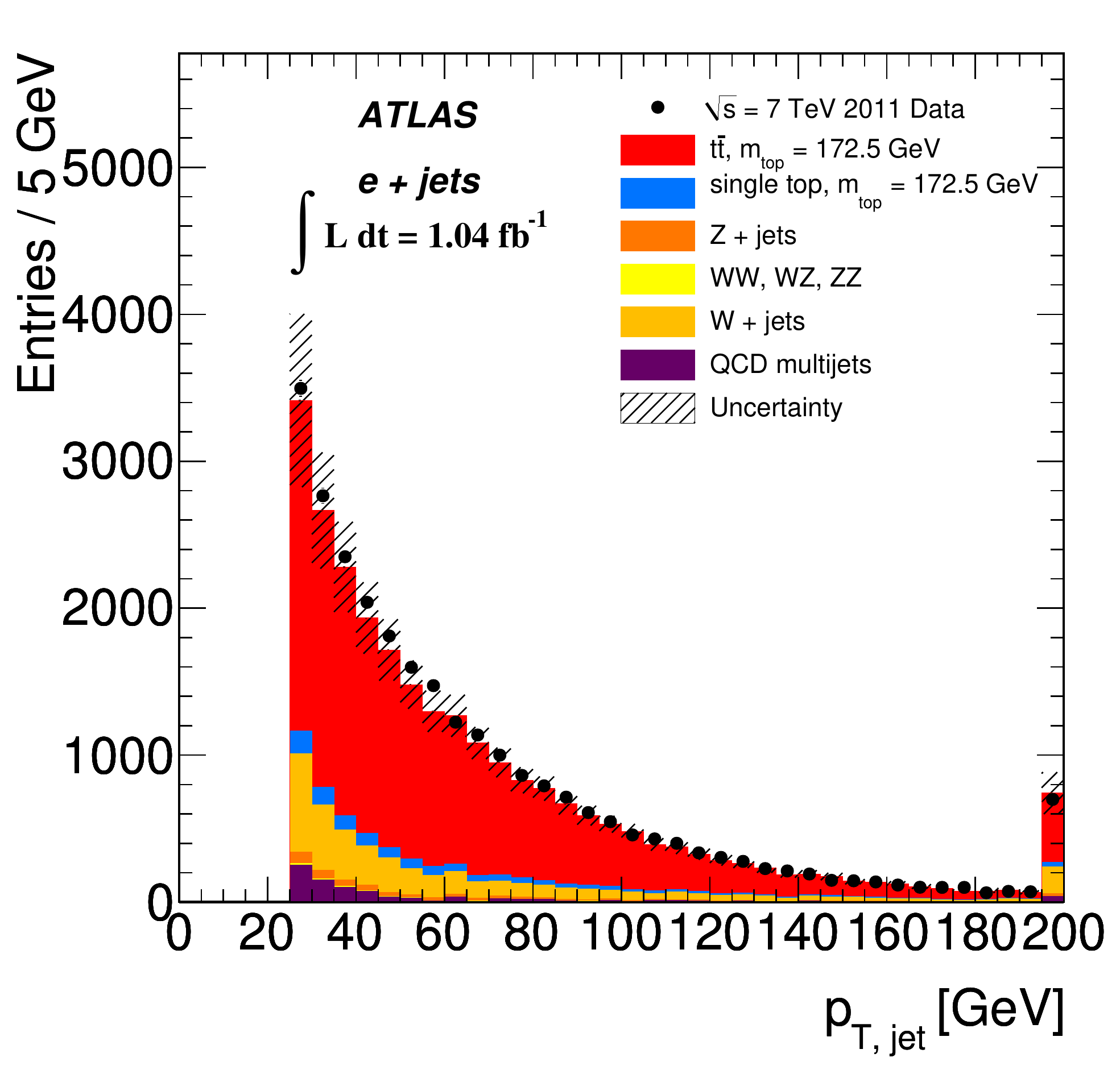}}
\subfigure[\mjets\ channel]{
  \includegraphics[width=0.38\textwidth]{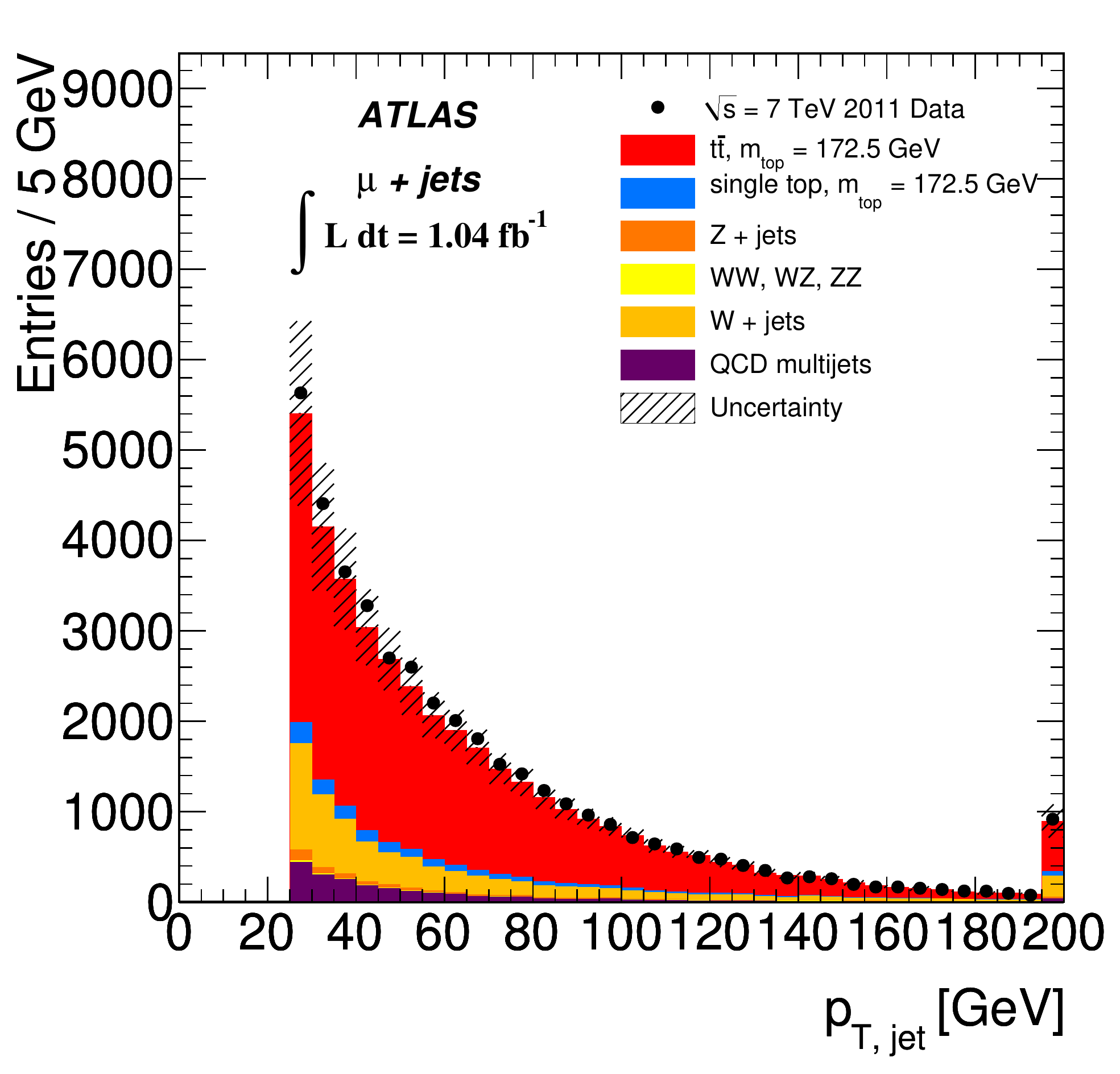}}
\subfigure[\ejets\ channel]{
  \includegraphics[width=0.38\textwidth]{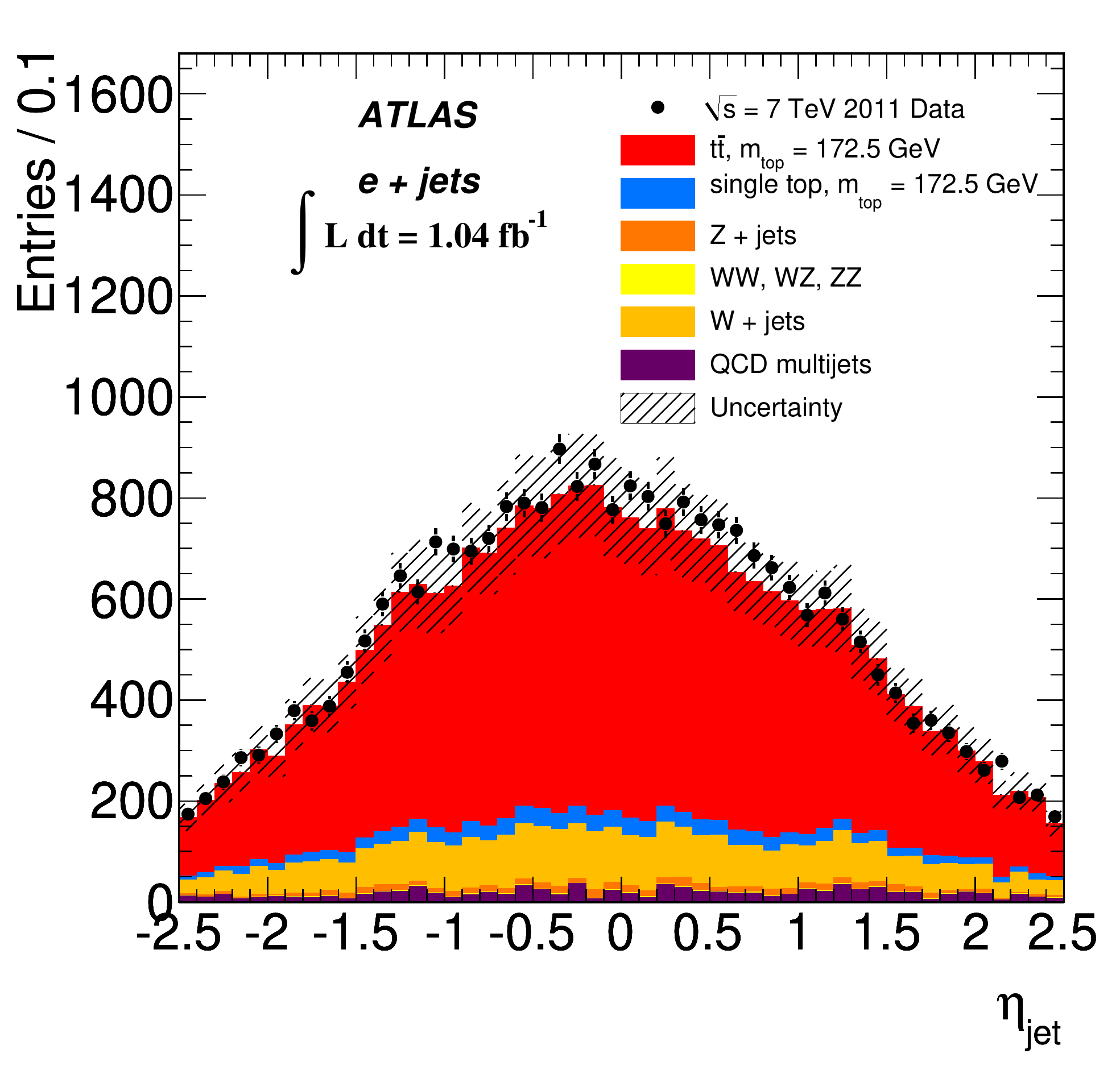}}
\subfigure[\mjets\ channel]{
  \includegraphics[width=0.38\textwidth]{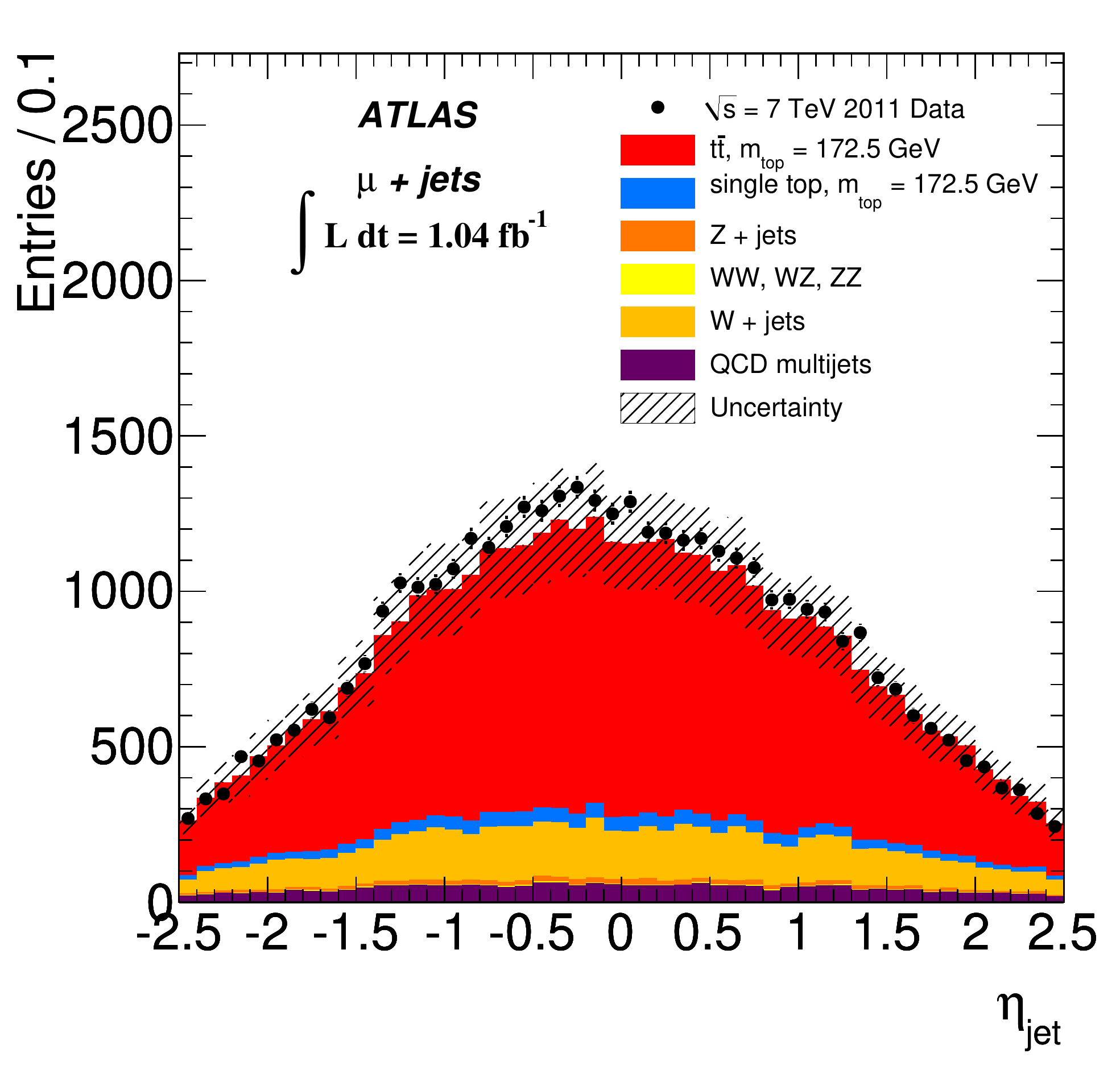}}
\caption{Distributions for the selected events of the common event selection in
  the \ejets\ channel on the left and the \mjets\ channel on the right. Shown
  are (a, b) the measured jet multiplicities, (c, d) the \pt, and (e, f) the
  $\eta$ distributions of all selected jets. The hatched area is the total
  uncertainty on the prediction described in the text. In (c, d) the rightmost
  bin also contains the overflow.
  \label{fig:tdjetkin}}
\end{figure*}
%

 For the \od, three additional requirements are applied. Firstly, only events
 with a converging likelihood fit (see Section~\ref{sec:1dim}) with a logarithm
 of the likelihood value $\lnL > \likecut$ are retained. Secondly, all jets in
 the jet triplet assigned to the hadronic decay of the top quark are required to
 fulfill $\pt>40$~\GeV, and thirdly the reconstructed \Wboson\ boson mass must
 lie within the range 60~\GeV~--~100~\GeV.

 For the \td\ the additional requirement is that only light jet pairs (see
 Section~\ref{sec:2dim}) with an invariant mass in the range
 50~\GeV~--~110~\GeV\ are considered for the \chiq\ fit.

 The numbers of events observed and expected, with the above selection and these
 additional analysis-specific requirements, are given in Table~\ref{tab:cutflow}
 for both channels and both analyses.
 For all Monte Carlo estimates, the uncertainties are the quadratic sum of the
 statistical uncertainty, the uncertainty on the \btag\ efficiencies, and a
 \atlumounc\ uncertainty on the luminosity~\cite{CON-2011-116,ATL-2011-002}.
 For the QCD multijet and the \Wj\ backgrounds, the systematic uncertainty
 estimated from data~\cite{CON-2011-023} dominates and is used instead.

 For both analyses and channels, the observed distributions for the leptons,
 jets, and kinematic properties of the top quark candidates such as their
 transverse momenta, are all well-described by the sum of the signal and
 background estimates.
 This is demonstrated for the properties of the selected jets, before applying
 the analysis specific requirements, for both channels in
 Figure~\ref{fig:tdjetkin}.
 The jet multiplicities, shown in Figure~\ref{fig:tdjetkin}(a, b), as well as
 the distributions of kinematic properties of jets like transverse momenta,
 Figure~\ref{fig:tdjetkin}(c, d), and the $\eta$ distributions,
 Figure~\ref{fig:tdjetkin}(e, f), are all well-described within the uncertainty
 band of the prediction.
 The size of the uncertainty band is dominated by the uncertainties on the
 background contributions estimated from data. The largest differences between
 the central values of the combined prediction and the data is observed for the
 rapidity distribution, with the data being higher, especially at central
 rapidities.
 Based on the selected events, the top quark mass is measured in two ways as
 described below.
%
\section{The 1d-analysis}
\label{sec:1dim}
%
 The \od\ is a one-dimensional template analysis using the reconstructed mass
 ratio:
%
\begin{eqnarray*}
  \RtW &\equiv& \frac{\mtr}{\mWr}.
\end{eqnarray*}
%
 Here \mtr\ and \mWr\ are the per event reconstructed invariant masses of the
 hadronically decaying top quark and \Wboson\ boson, respectively.

 To select the jet triplet for determining the two masses, this analysis
 utilises a kinematic fit maximising an event likelihood.
 This likelihood relates the observed objects to the \ttbar\ decay products
 (quarks and leptons) predicted by the NLO signal Monte Carlo, albeit in a
 Leading Order (LO) kinematic approach, using
 $\ttbar\to\ell\nu\blep\,\qone\qtwo\bhad$.
 In this procedure, the measured jets relate to the quark decay products of the
 \Wboson\ boson, \qone\ and \qtwo, and to the \bquarks, \blep\ and \bhad,
 produced in the top quark decays.
 The \met\ vector is identified with the transverse momentum components of the
 neutrino, $\hat{p}_{x,\nu}$ and $\hat{p}_{y,\nu}$.

 The likelihood is defined as a product of transfer functions (\tf),
 Breit-Wigner (\bw) distributions, and a weight \Wbtag\ accounting for the
 \btag\ information:
%
\begin{figure}[tbp!]
\centering
\subfigure[\ejets\ channel]{
  \includegraphics[width=0.44\textwidth]{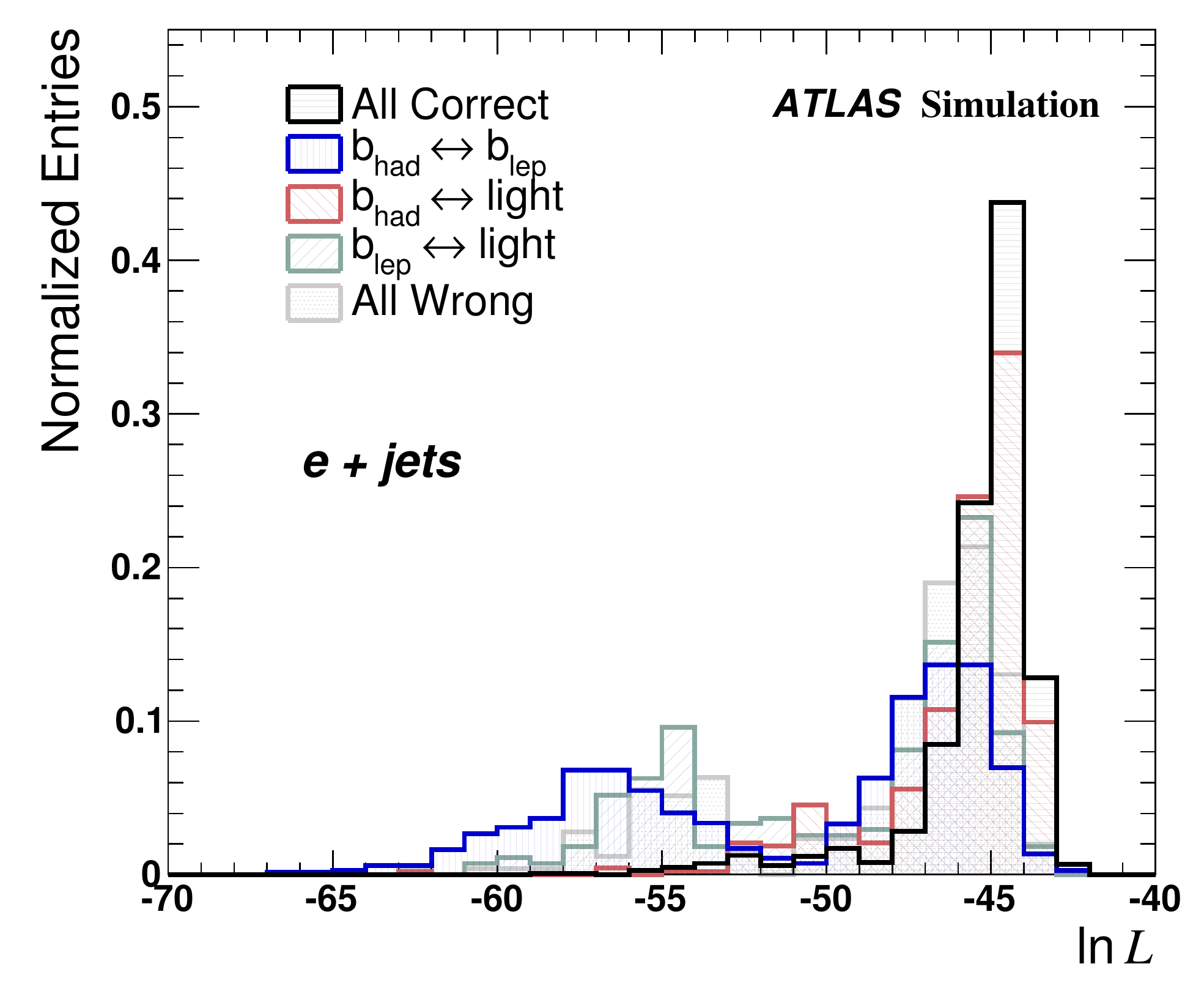}}
\subfigure[\ejets\ channel]{
  \includegraphics[width=0.44\textwidth]{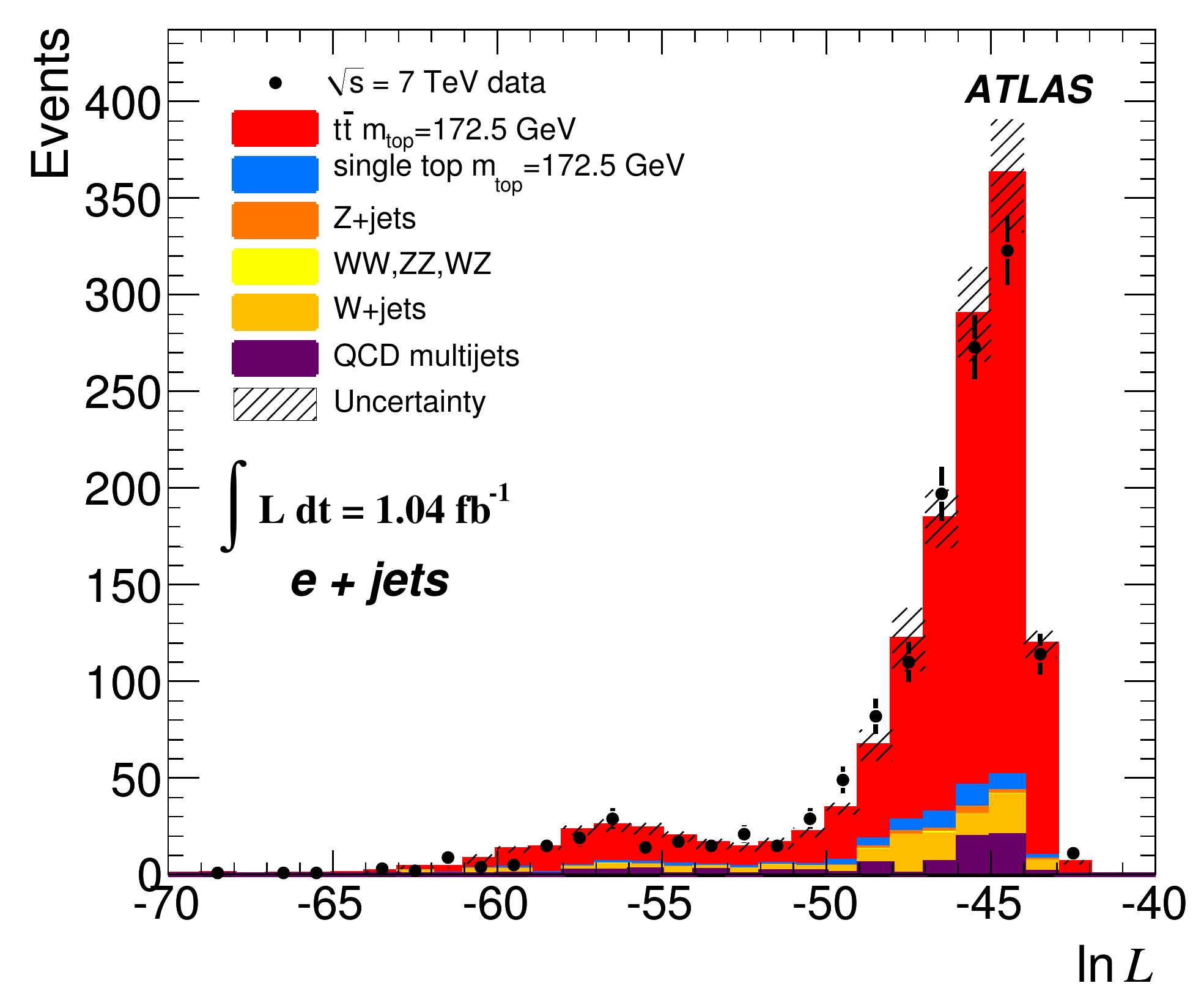}}
\subfigure[\ejets\ channel]{
  \includegraphics[width=0.44\textwidth]{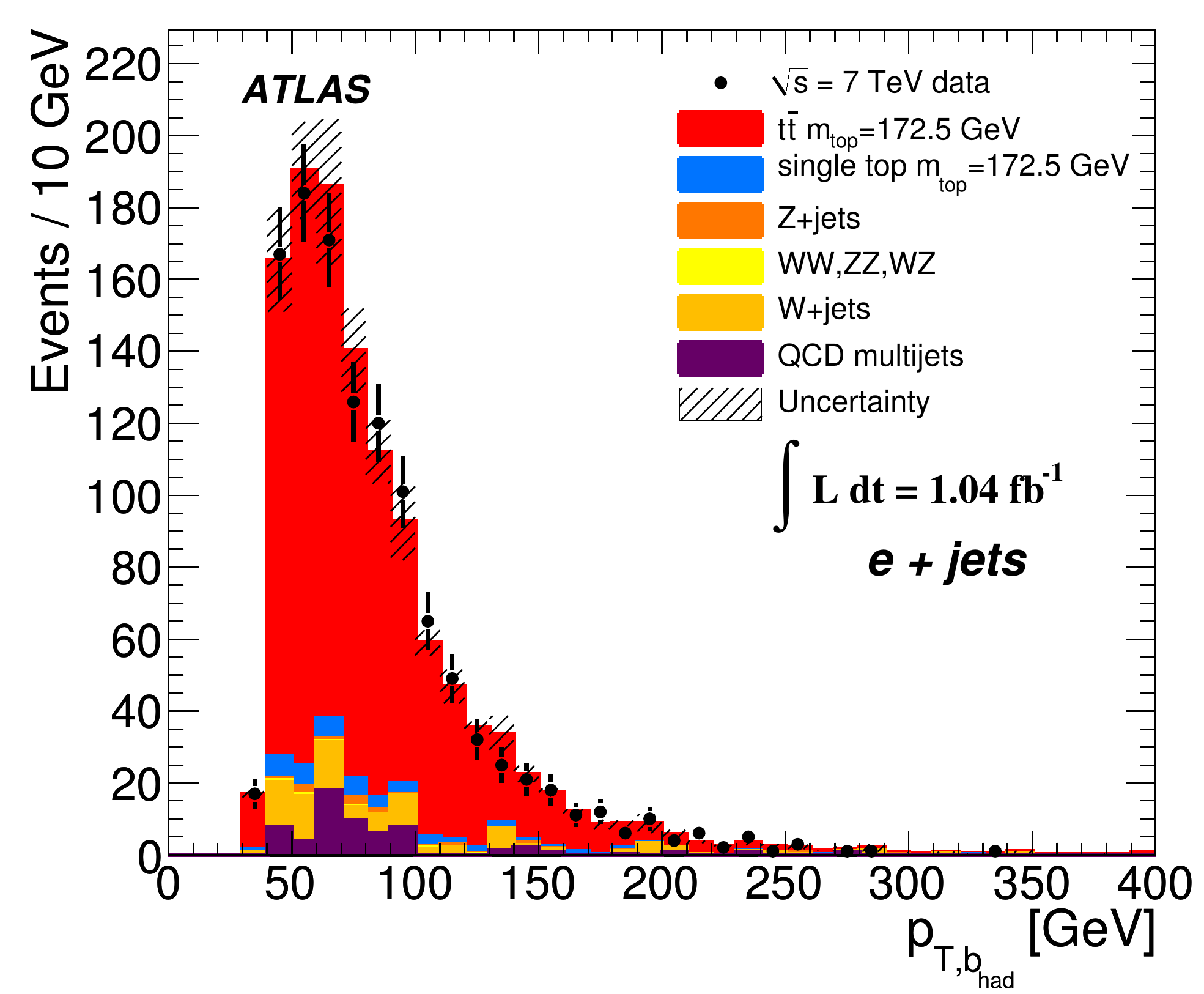}}
\caption{{\bf \od:} Performance of the likelihood fit in the
  \ejets\ channel. Shown in (a) are the predicted \lnL\ distributions for
  various jet permutations in the \ttbar\ signal Monte Carlo. The figures (b, c)
  compare two output variables of the likelihood fit as observed in the data
  with their respective prediction. These are (b) the \lnL\ value, and (c) the
  \pt\ of the \bjet\ associated to the hadronic decay of the top quark.
  \label{fig:odlikeliel}}
\end{figure}
%
%
\begin{eqnarray*}
 L & = &
\tf\left(\Ejet{1} \vert \hat{E}_{\bhad}\right) \cdot
\tf\left(\Ejet{2} \vert \hat{E}_{\blep}\right) \cdot
\tf\left(\Ejet{3} \vert \hat{E}_{\qone}\right) \cdot
 \\ &&
\tf\left(\Ejet{4} \vert \hat{E}_{\qtwo}\right) \cdot
\tf\left(\Emis{x} \vert \hat{p}_{x,\nu}\right) \cdot
\tf\left(\Emis{y} \vert \hat{p}_{y,\nu}\right) \cdot
\nonumber \\ &&
\left\{ \begin{array}{cr}
\tf\left(E_{e}      \vert \hat{E}_{e} \right) & \ejets \\ 
\tf\left(p_{\mathrm{T,\mu}} \vert \hat{p}_{\mathrm{T},\mu} \right) & \mjets 
\end{array} \right\}  \cdot 
\nonumber \\ &&
\bw\left[m(\qone\, \qtwo) \vert \mW, \GW\right] \cdot
\bw\left[m(\ell\,\nu) \vert \mW, \GW\right] \cdot 
\nonumber \\ &&
\bw\left[m(\qone\,\qtwo\,\bhad) \vert \mtrl, \Gt \right] \cdot
\nonumber \\ &&
\bw\left[m(\ell\,\nu\,\blep)\vert \mtrl, \Gt \right]\cdot\Wbtag\,.
\nonumber 
\label{eq:kinlike}
\end{eqnarray*}
%
 \begin{sloppypar}
 The generator predicted quantities are marked with a circumflex
 (e.g.~$\hat{E}_{\bhad}$), i.e.~the energy of the \bquark\ from the hadronic
 decay of the top quark.
 The quantities \mW\ and \GW\ (which amounts to about one fifth of the Gaussian
 resolution of the \mWr\ distribution) are taken from Ref.~\cite{NAK-1001}, and
 \mtrl\ is the likelihood estimator for the top quark mass, i.e.~the per event
 result of maximising this likelihood.
 Transfer functions are derived from the \Mcatnlo\ \ttbar\ signal Monte Carlo
 sample at an input mass of $\mt=172.5$~\GeV, based on reconstructed objects
 that are matched to their generator predicted quarks and leptons.
 When using a maximum separation of $\Delta R=0.4$ between a quark and the
 corresponding jet, the fraction of events with four matched jets from all
 selected events amounts to \fullmalow~--~\fullmahig.
 The transfer functions are obtained in three bins of \eta\ for the energies of
 \bquark\ jets, $\Ejet{1}$ and $\Ejet{2}$, light quark jets, $\Ejet{3}$ and
 $\Ejet{4}$, the energy, $E_{e}$, (or transverse momentum,
 $p_{\mathrm{T,\mu}}$) of the charged lepton, and the two components of the
 \met, $\Emis{x}$ and $\Emis{y}$.
 In addition, the likelihood exploits the values of \mW\ and \GW\ to constrain
 the reconstructed leptonic, $m(\ell\,\nu)$, and hadronic, $m(\qone\, \qtwo)$,
 \Wboson\ boson masses using Breit-Wigner distributions.
%
\begin{figure}[tbp!]
\centering
\subfigure[\ejets\ channel]{
  \includegraphics[width=0.47\textwidth]{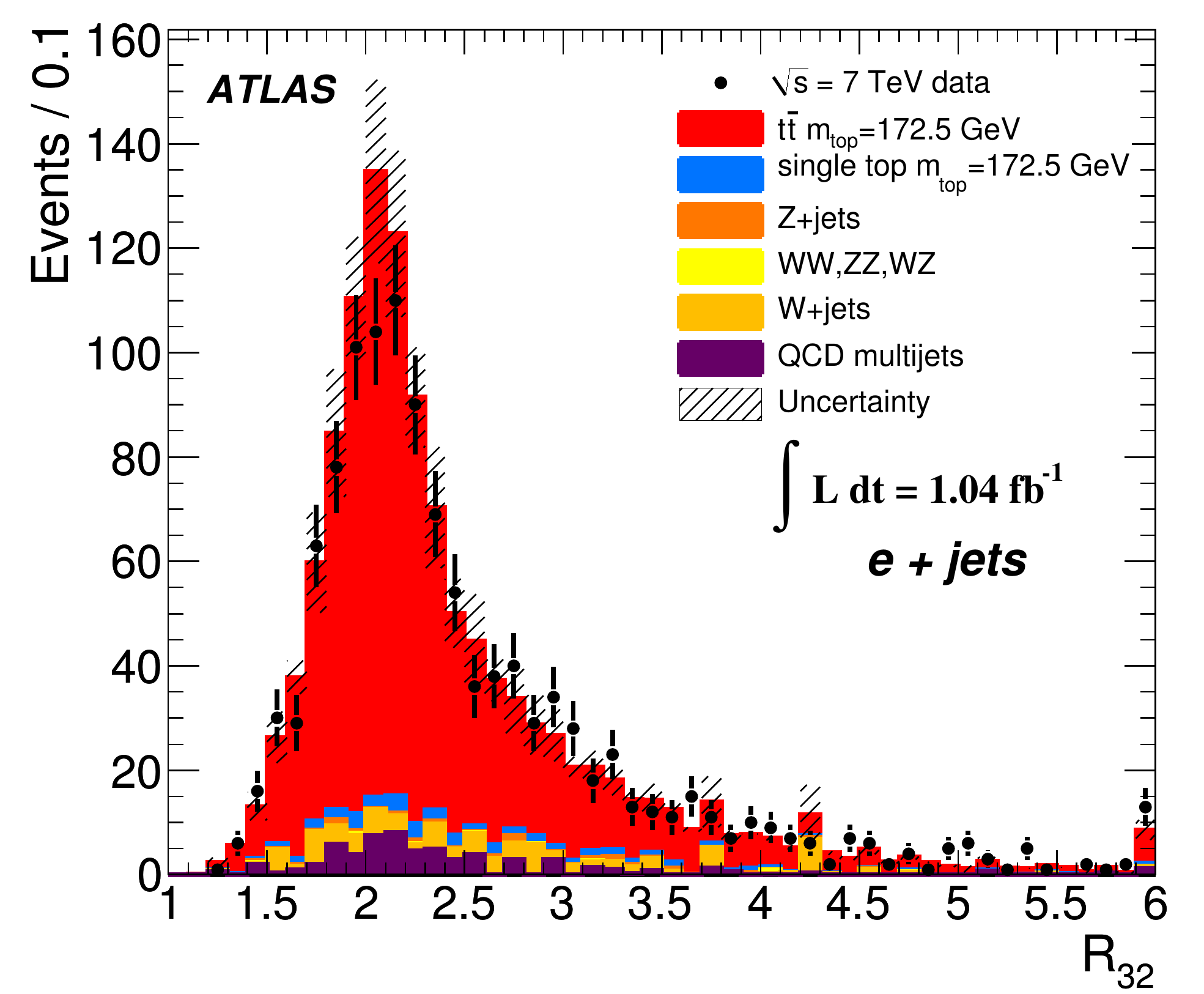}}
\subfigure[\mjets\ channel]{
  \includegraphics[width=0.47\textwidth]{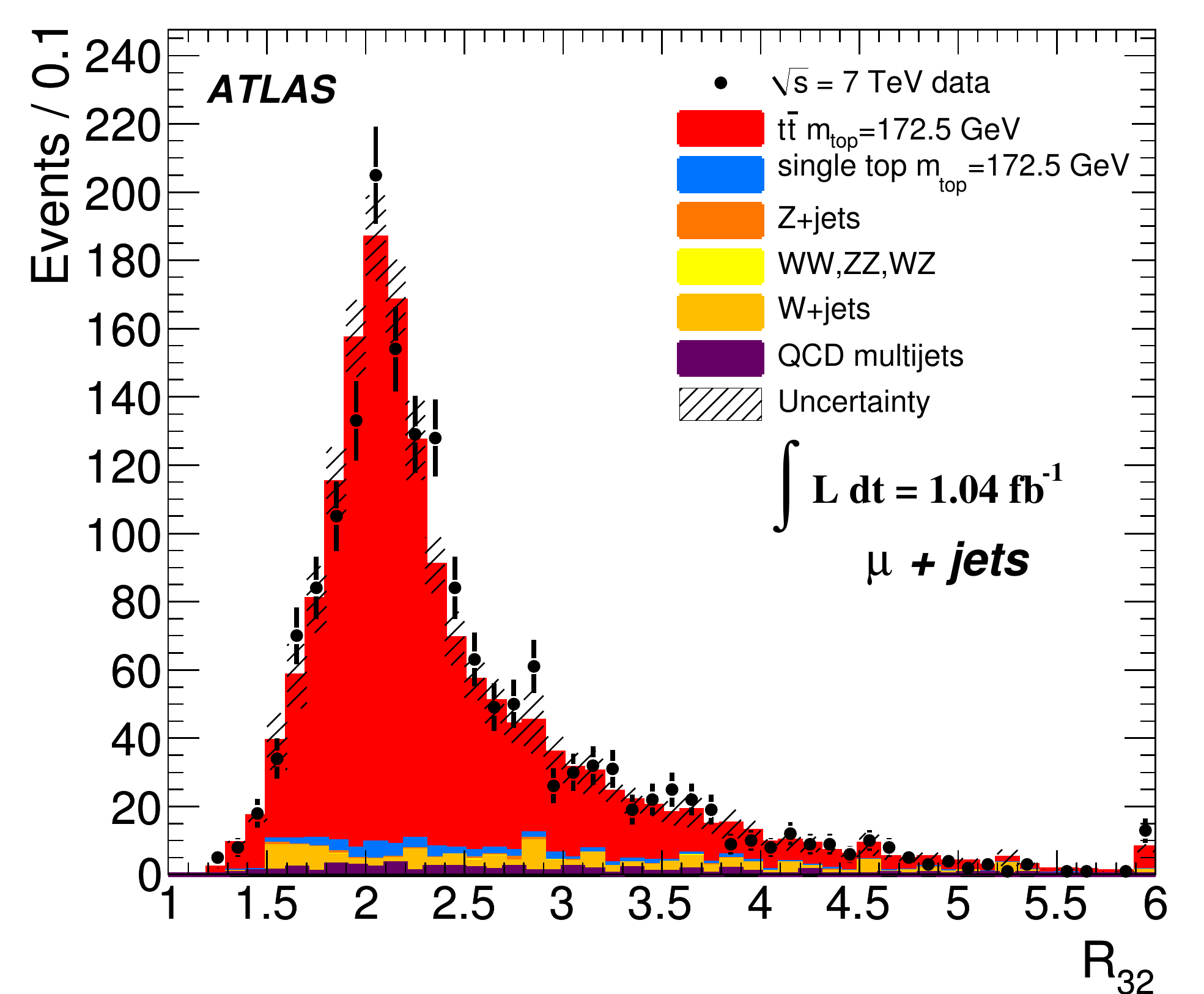}}
\caption{{\bf \od:} The reconstructed \RtW\ constructed from the selected jet
  permutation using the unconstrained four-vectors of the jet triplet for (a)
  the \ejets\ channel, and (b) the \mjets\ channel. The rightmost bins also
  contain the overflow.
  \label{fig:odratio}}
\end{figure}
%
 Similarly, the reconstructed leptonic, $m(\ell\,\nu\,\blep)$, and hadronic,
 $m(\qone\,\qtwo\,\bhad)$, top quark masses are constrained to be identical,
 where the width of the corresponding Breit Wigner distribution is identified
 with the predicted \Gt (using its top quark mass dependence)~\cite{NAK-1001}.
 Including the \btag\ information into the likelihood as a weight \Wbtag,
 derived from the efficiency and mistag rate of the \btag\ algorithm, and
 assigned per jet permutation according to the role of each jet for a given jet
 permutation, improves the selection of the correct jet permutation.
 As an example, for a permutation with two \bjets\ assigned to the
 \bquark\ positions and two light jets to the light quark positions, the weight
 \Wbtag\ amounts to 0.48, i.e.~it corresponds to the square of the
 \btag\ efficiency times the square of one minus the fake rate, both given
 in Section~\ref{sec:evsel}.
 \end{sloppypar}

 With this procedure, the correct jet triplet for the hadronic top quark is
 chosen in about \jetcorper\ of simulated signal events with four matched jets.
 However, if \RtW\ from the likelihood fit, i.e.~calculated from \mtrl\ and
 \mWrl, is taken, a large residual jet energy scale (JES) dependence of \RtW
 remains.
 This is because in the fit \mWr\ is constrained to \mW, while \mtr\ is only
 constrained to be equal for the leptonic and hadronic decays of the top quarks.
 This spoils the desired event-by-event reduction of the JES uncertainty in the
 ratio \RtW~\cite{CON-2011-033}.
 To make best use of the high selection efficiency for the correct jet
 permutation from the likelihood fit, and the stabilisation of \RtW\ against JES
 variations, the jet permutation derived in the fit is used, but \mWr, \mtr\ and
 therefore \RtW, are constructed from the unconstrained four-vectors of the jet
 triplet as given by the jet reconstruction.

 The performance of the algorithm, shown in Figure~\ref{fig:odlikeliel} for the
 \ejets\ channel, is similar for both channels.
 The likelihood values of wrong jet permutations for signal events from the
 large \Mcatnlo\ sample are frequently considerably lower than the ones for the
 correct jet permutations, as seen in Figure~\ref{fig:odlikeliel}(a).
 For example, the distribution for the jet permutation in which the jet from the
 \bquark\ from the leptonically decaying top quark is exchanged with one light
 quark jet from the hadronic \Wboson\ boson decay has a second peak at about ten
 units lower than the one for the correct jet permutation.
 The actual distribution of \lnL\ values observed in the data is well-described
 by the signal plus background predictions, as seen in
 Figure~\ref{fig:odlikeliel}(b).
 The kinematic distributions of the variables used in the transfer functions are
 also well-described by the predictions, as shown in
 Figure~\ref{fig:odlikeliel}(c), for the example of the resulting \pt\ of the
 \bjet\ associated to the hadronic decay of the top quark.
 The resulting \RtW\ distributions for both channels are shown in
 Figure~\ref{fig:odratio}. They are also well accounted for by the predictions.

 Signal templates are derived for the \RtW\ distribution for all \mt\ dependent
 samples, consisting of the \ttbar\ signal events, together with single top
 quark production events.
 This procedure is adopted, firstly, because single top quark production,
 although formally a background process, still carries information about the top
 quark mass and, secondly, by doing so \mt\ independent background templates can
 be used.
 The templates are constructed for the six \mt\ choices using the specifically
 generated Monte Carlo samples, see Section~\ref{sec:mc}.

 The \RtW\ templates are parameterised with a functional form given by the sum
 of a ratio of two correlated Gaussians and a Landau function.
 The ratio of two Gaussians~\cite{HIN-6901} is motivated as a representation of
 the ratio of two correlated measured masses.
 The Landau function is used to describe the tails of the distribution stemming
 mainly from wrong jet-triplet assignments.
 The correlation between the two Gaussian distributions is fixed to 50$\%$.
 A simultaneous fit to all templates per decay channel is used to derive a
 continuous function of \mt\ that interpolates the \RtW\ shape differences among
 all mass points with \mt\ in the range described above.
 This approach rests on the assumption that each parameter has a linear
 dependence on the top quark mass, which has been verified for both channels.
 The fit minimises a \chiq\ built from the \RtW\ distributions at all mass
 points simultaneously.
 The \chiq\ is the sum over all bins of the difference squared between the
 template and the functional form, divided by the statistical uncertainty
 squared in the template.
 The combined fit adequately describes the \RtW\ distributions for both
 channels.
 In Figure~\ref{fig:odtemplel}(a) the sensitivity to \mt\ is shown in the
 \ejets\ channel by the superposition of the signal templates and their fits for
 four of the six input top quark masses assumed in the simulation.
%
\begin{figure}[tbp!]
\centering
\subfigure[\ejets\ channel]{
  \includegraphics[width=0.47\textwidth]{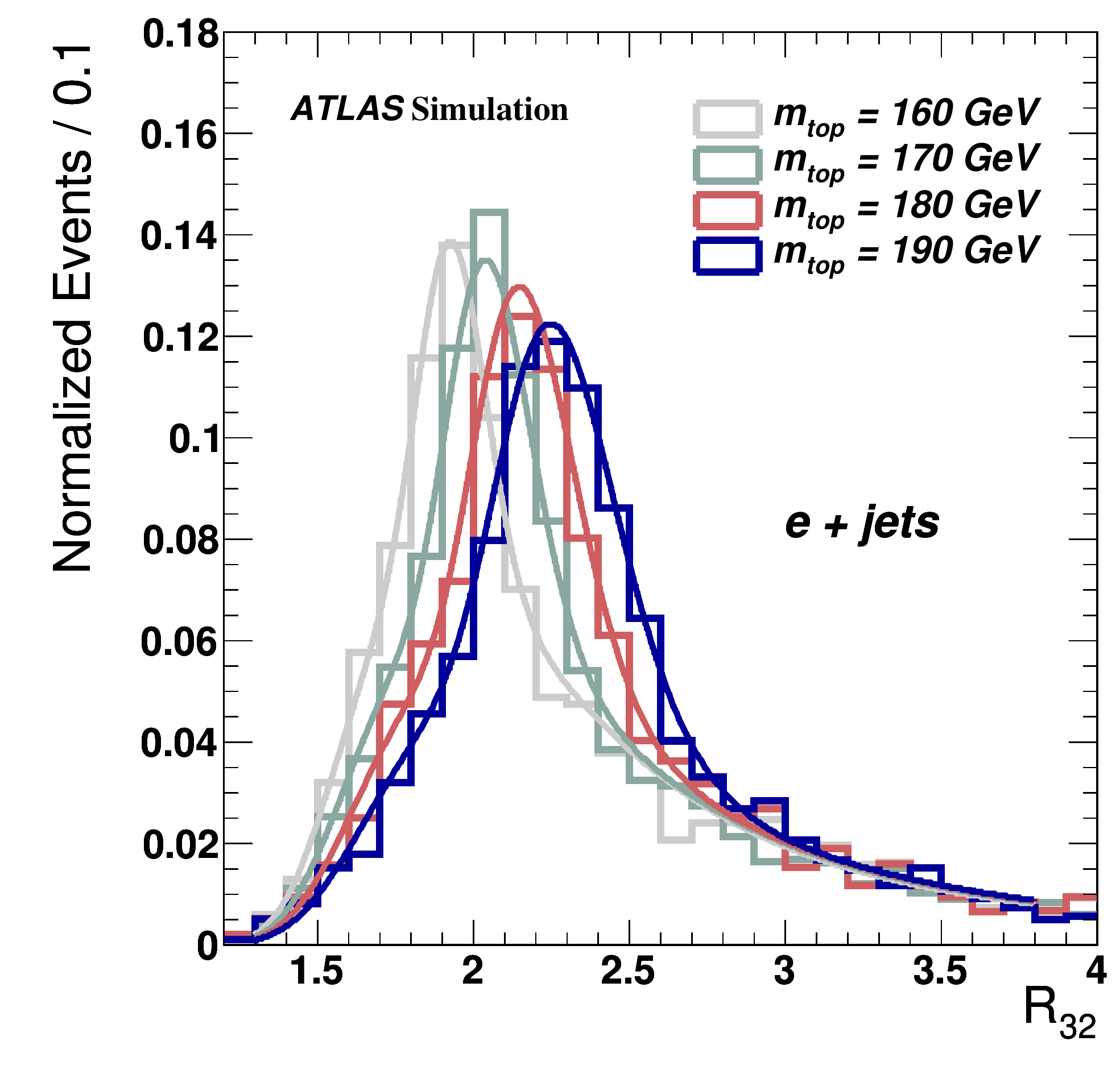}}
\subfigure[\ejets\ channel]{
  \includegraphics[width=0.47\textwidth]{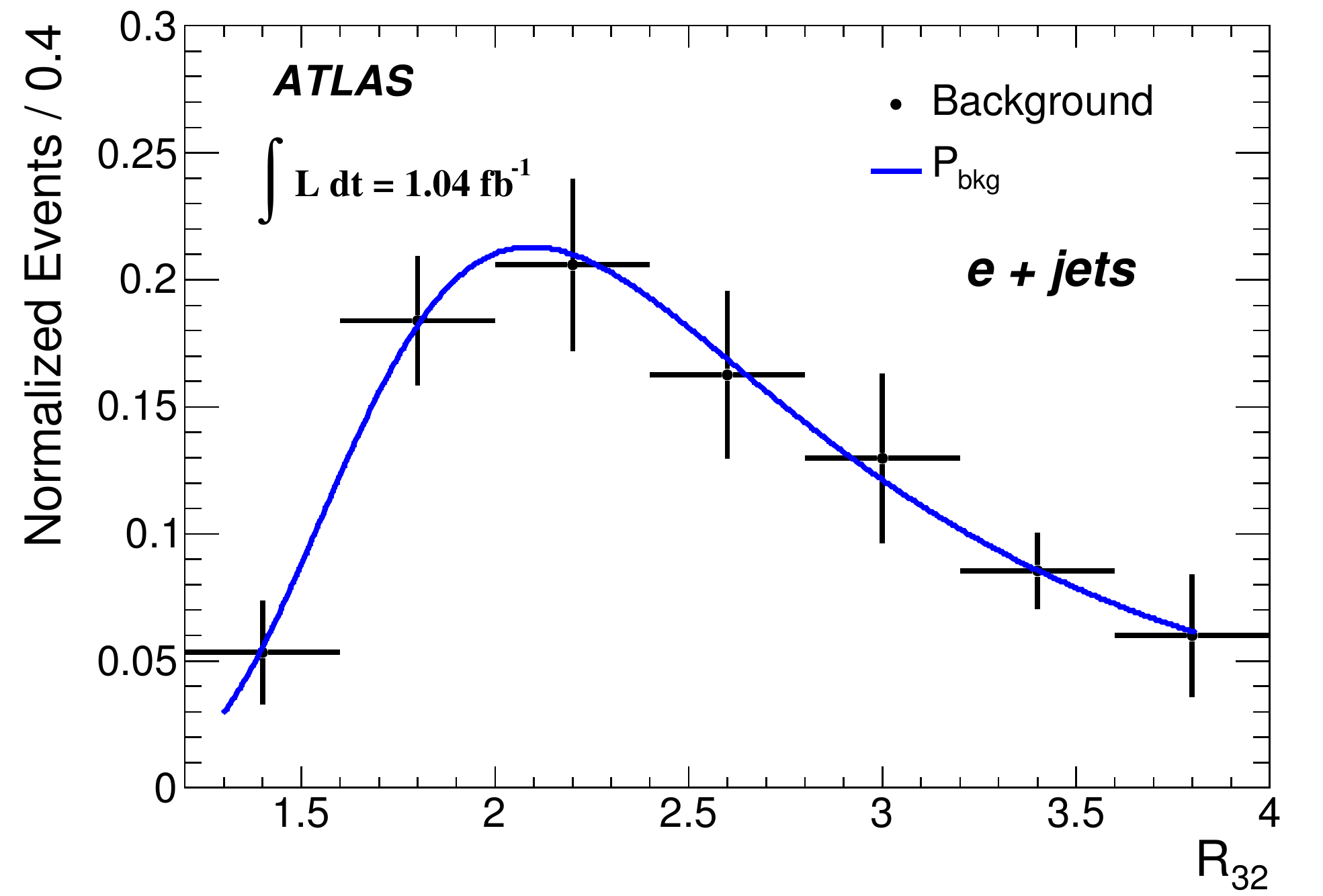}}
\caption{{\bf \od:} Template parameterisations for (a) signal and (b) background
  contributions in the \ejets\ channel. The background fit is labelled \Pbkg.
  \label{fig:odtemplel}}
\end{figure}
%

 For the background template, the \mt\ independent parts, see
 Table~\ref{tab:cutflow}, are treated together.
 Their individual distributions, taken either from Monte Carlo or data estimates
 as detailed above, are summed, and a Landau distribution is chosen to
 parameterise their \RtW\ distribution.
 For each channel this function adequately describes the background distribution
 as shown in Figure~\ref{fig:odtemplel}(b) for the \ejets\ channel, which has a
 larger background contribution than the \mjets\ channel.

 Signal and background probability density functions, $\Psig(\RtW\vert\mt)$ and
 $\Pbkg(\RtW)$, respectively, are used in a binned likelihood fit to the data
 using a number of bins, \Nbins. The likelihood reads:
%
\begin{eqnarray*}
 \Like{}(\RtW\vert\mt)      &=& \Like{shape}(\RtW\vert\mt)\times\Like{bkg}(\RtW)\,,\\
 \Like{shape}(\RtW\vert\mt) &=& \prod_{i=1}^{\Nbins}
                            \frac{\lami^{\Ni}}{\Ni!}\cdot e^{-\lami}\,,\nonumber\\
 \Like{bkg}(\RtW)           &=& \exp{\left\{-\frac{(\nbkg-\nbpred)^2}{2\sigma^2_{\nbpred}}\right\}}
                            \,\,,\nonumber
\end{eqnarray*}
\noindent
with:
\begin{eqnarray*}
      \lami &=& (N-\nbkg)\cdot\Psig(\RtW\vert\mt)_{i}+\nbkg\cdot\Pbkg(\RtW)_{i}\,\,,\nonumber\\
          N &=& \sum_{i=1}^{\Nbins}\Ni = \nsig+\nbkg\,\,.\nonumber
\end{eqnarray*}
%
 The variable \Ni\ denotes the number of events observed per bin, and \nsig\ and
 \nbkg\ denote the total numbers of signal and background events to be
 determined.
 The term \Like{shape} accounts for the shape of the \RtW\ distribution and its
 dependence on the top quark mass \mt.
 The term \Like{bkg} constrains the total number of background events, \nbkg,
 using its prediction, \nbpred, and the background uncertainty, chosen to be
 \buncpro, see Table~\ref{tab:cutflow}. In addition, the number of background
 events is restricted to be positive.
 The two free parameters of the fit are the total number of background events,
 \nbkg, and \mt.
 The performance of this algorithm is assessed with the pseudo-experiment
 technique.
 For each \mt\ value, distributions from pseudo-experiments are constructed by
 random sampling of the simulated signal and background events used to construct
 the corresponding templates.
 Using Poisson statistics, the numbers of signal events and total background
 events in each pseudo-experiment are fluctuated around the expectation values,
 either calculated assuming SM cross-sections and the integrated luminosity of
 the data, or taken from the data estimate.
 A good linearity is found between the input top quark mass used to perform the
 pseudo-experiments, and the result of the fit.
 Within their statistical uncertainties, the mean values and width of the pull
 distributions are consistent with the expectations of zero and one,
 respectively.
 The expected statistical uncertainties (mean $\pm$ RMS) obtained from
 pseudo-experiments with an input top quark mass of $\mt=172.5$~\GeV, and for a
 luminosity of 1~\ifb, are \mtstaelod~\GeV\ and \mtstamuod~\GeV\ for the
 \ejets\ and \mjets\ channels, respectively.
%
\section{The 2d-analysis}
\label{sec:2dim}
%
 \begin{sloppypar}
 In the \td, similarly to Ref.~\cite{CDF-0601}, \mt\ and a global jet energy
 scale factor (JSF) are determined simultaneously by using the \mtr\ and
 \mWr\ distributions\footnote{Although for the two analyses \mtr\ and \mWr\ are
   calculated differently, the same symbols are used to indicate that these are
   estimates of the same quantities.}.
 Instead of stabilising the estimator of \mt\ against JES variations as done for
 the \od, the emphasis here is on an in-situ jet scaling.
 A global JSF (averaged over $\eta$ and \pt) is obtained, which is mainly based
 on the observed differences between the predicted \mWr\ distribution and the
 one observed for the data.
 This algorithm predicts which global JSF correction should be applied to all
 jets to best fit the data.
 Due to this procedure, the JSF is sensitive not only to the JES, but also to
 all possible differences in data and predictions from specific assumptions made
 in the simulation that can lead to differences in the observed jets. 
 These comprise: the fragmentation model, initial state and final state QCD
 radiation (ISR and FSR), the underlying event, and also pileup.
%
\begin{figure*}[tbp!]
\centering
\subfigure[\ejets\ channel]{
  \includegraphics[width=0.47\textwidth]{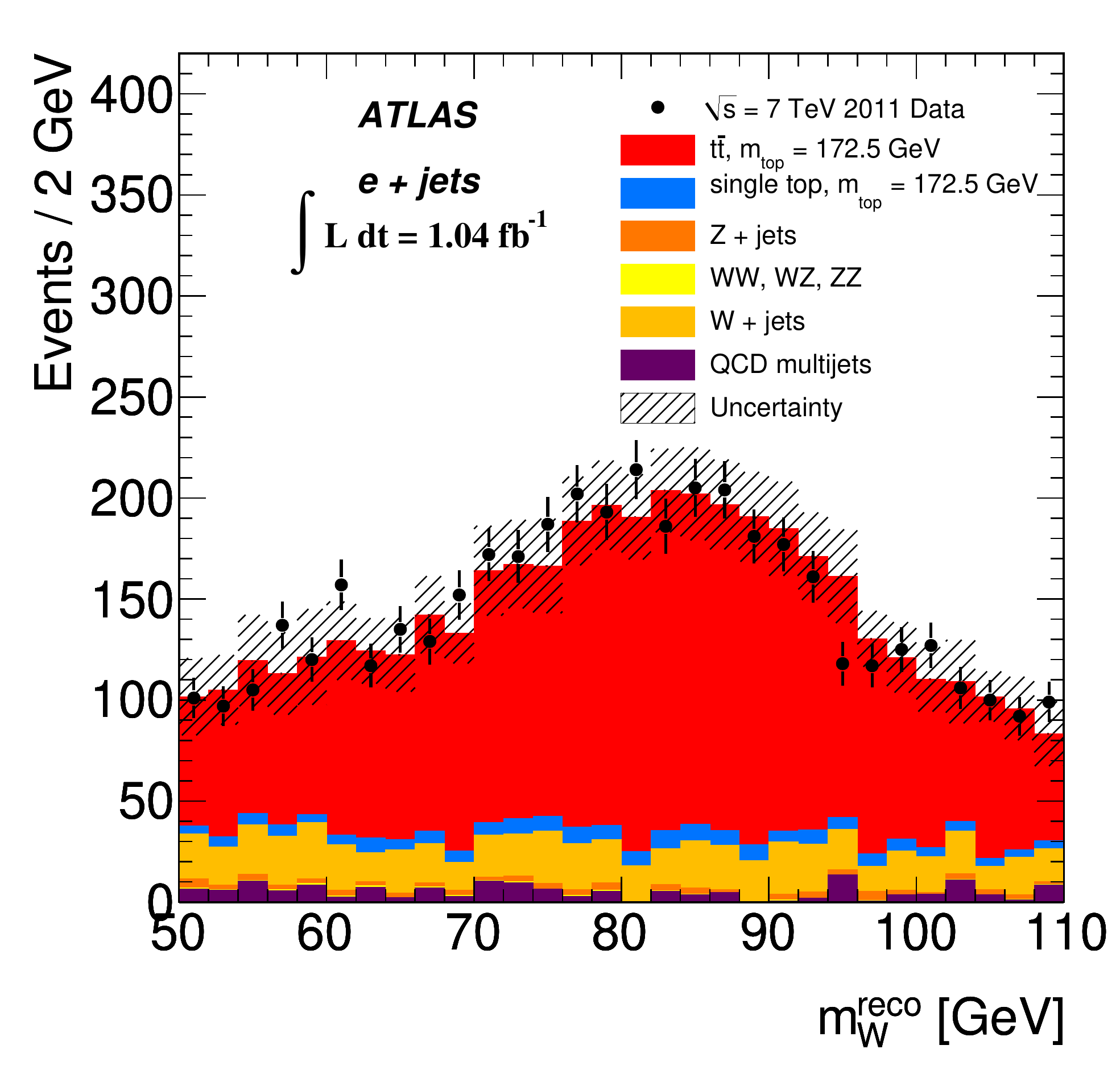}}
\subfigure[\mjets\ channel]{
  \includegraphics[width=0.47\textwidth]{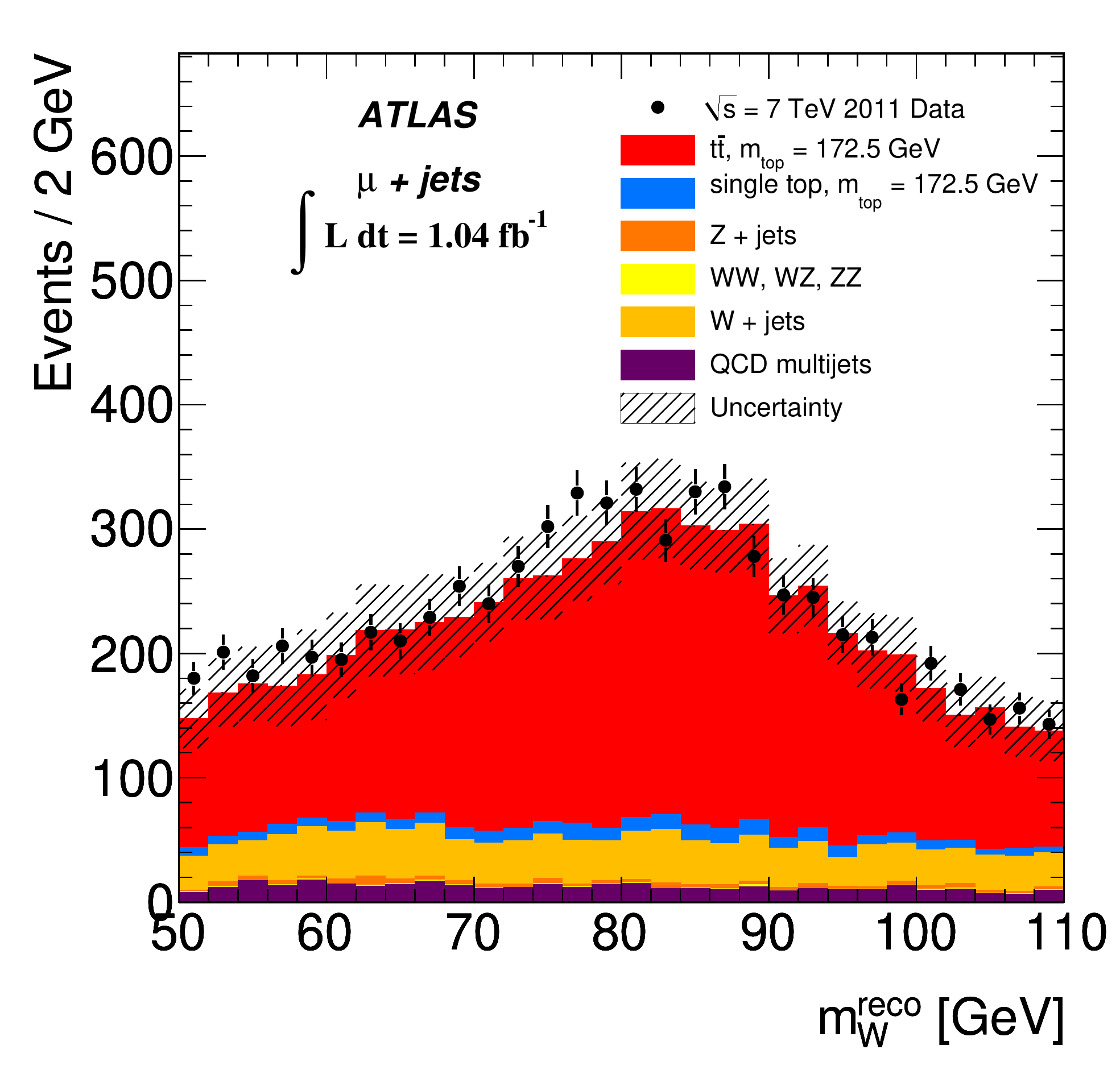}}
\subfigure[\ejets\ channel]{
  \includegraphics[width=0.47\textwidth]{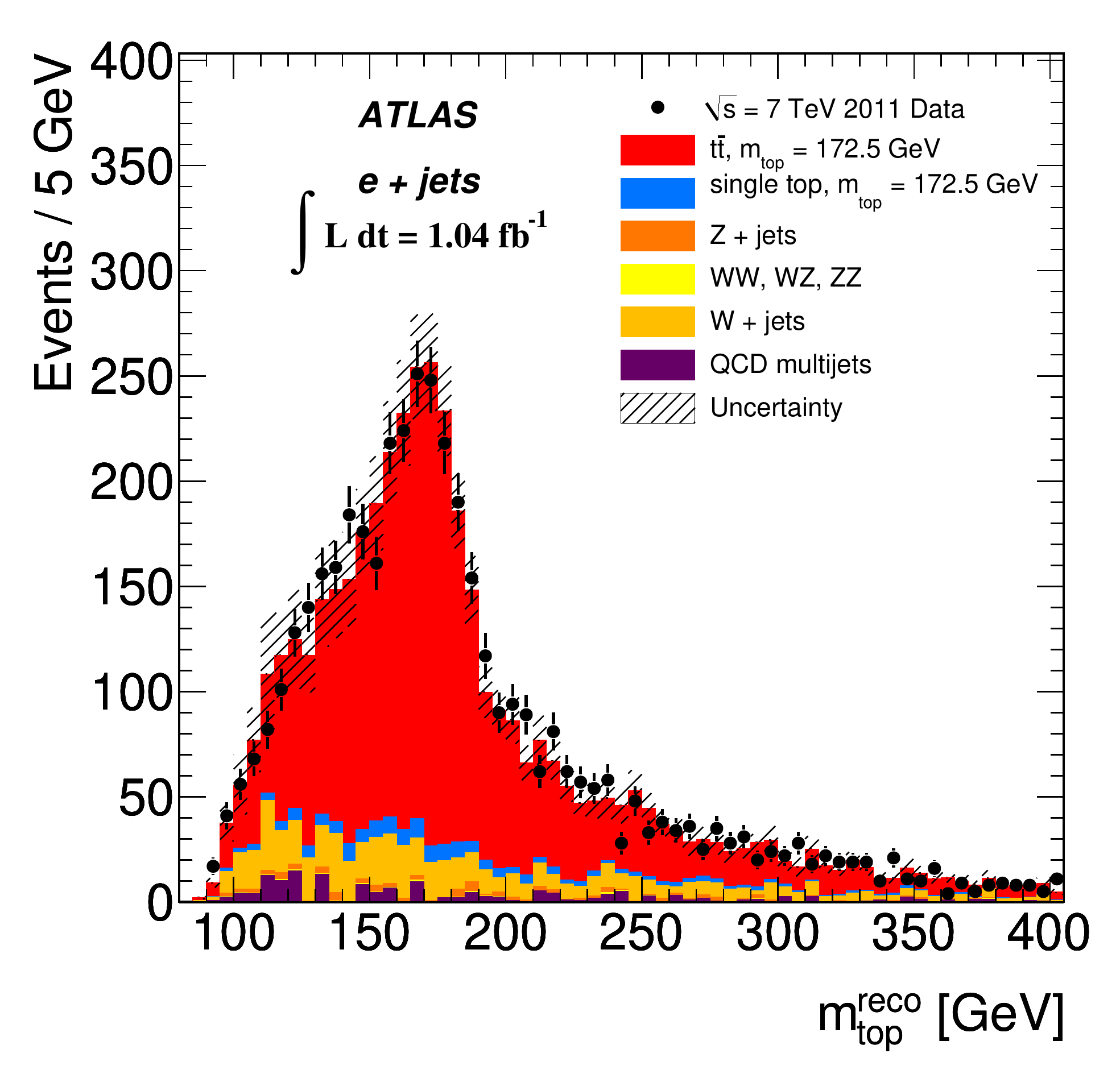}}
\subfigure[\mjets\ channel]{
  \includegraphics[width=0.47\textwidth]{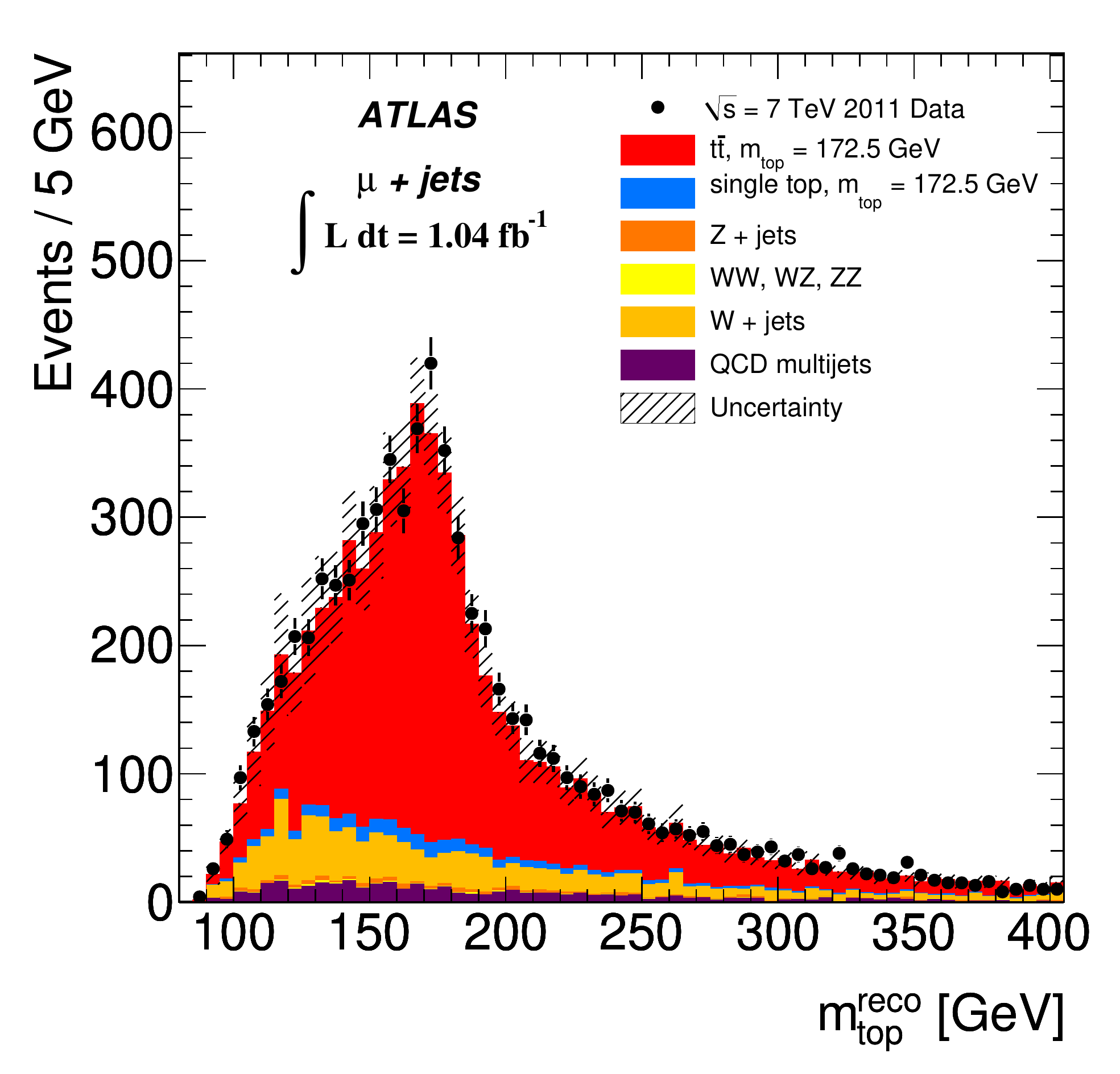}}
\caption{{\bf \td:} Reconstructed \Wboson\ boson and top quark masses, \mWr\ and
  \mtr, observed in the data together with the signal and background
  predictions. Shown are (a, c) the \ejets\ channel, and (b, d) the
  \mjets\ channel.
  \label{fig:tdmass}}
\end{figure*}
%
 In this method, the systematic uncertainty on \mt\ stemming from the JES is
 reduced and partly transformed into an additional statistical uncertainty on
 \mt\ due to the two-dimensional fit.
 The precisely measured values of \mW\ and \GW~\cite{NAK-1001} are used to
 improve on the experimental resolution of \mtr\ by relating the observed jet
 energies to the corresponding parton energies as predicted by the signal Monte
 Carlo (i.e.~to the two quarks from the hadronic \Wboson\ boson decay, again
 using LO kinematics).
 Thereby, this method offers a complementary determination of \mt\ to the
 \od\ method, described in Section~\ref{sec:1dim}, with different sensitivity to
 systematic effects and data statistics.
 \end{sloppypar}

 \begin{sloppypar}
 For the events fulfilling the common requirements listed in
 Section~\ref{sec:evsel}, the jet triplet assigned to the hadronic top quark
 decay is constructed from any \bjet, together with any light jet pair with a
 reconstructed \mWr\ within 50~\GeV~--~110~\GeV.
 Amongst those, the jet triplet with maximum \pt\ is chosen as the top quark
 candidate.
%
\begin{figure*}[tbp!]
\centering
\subfigure[\mjets\ channel]{
  \includegraphics[width=0.47\textwidth]{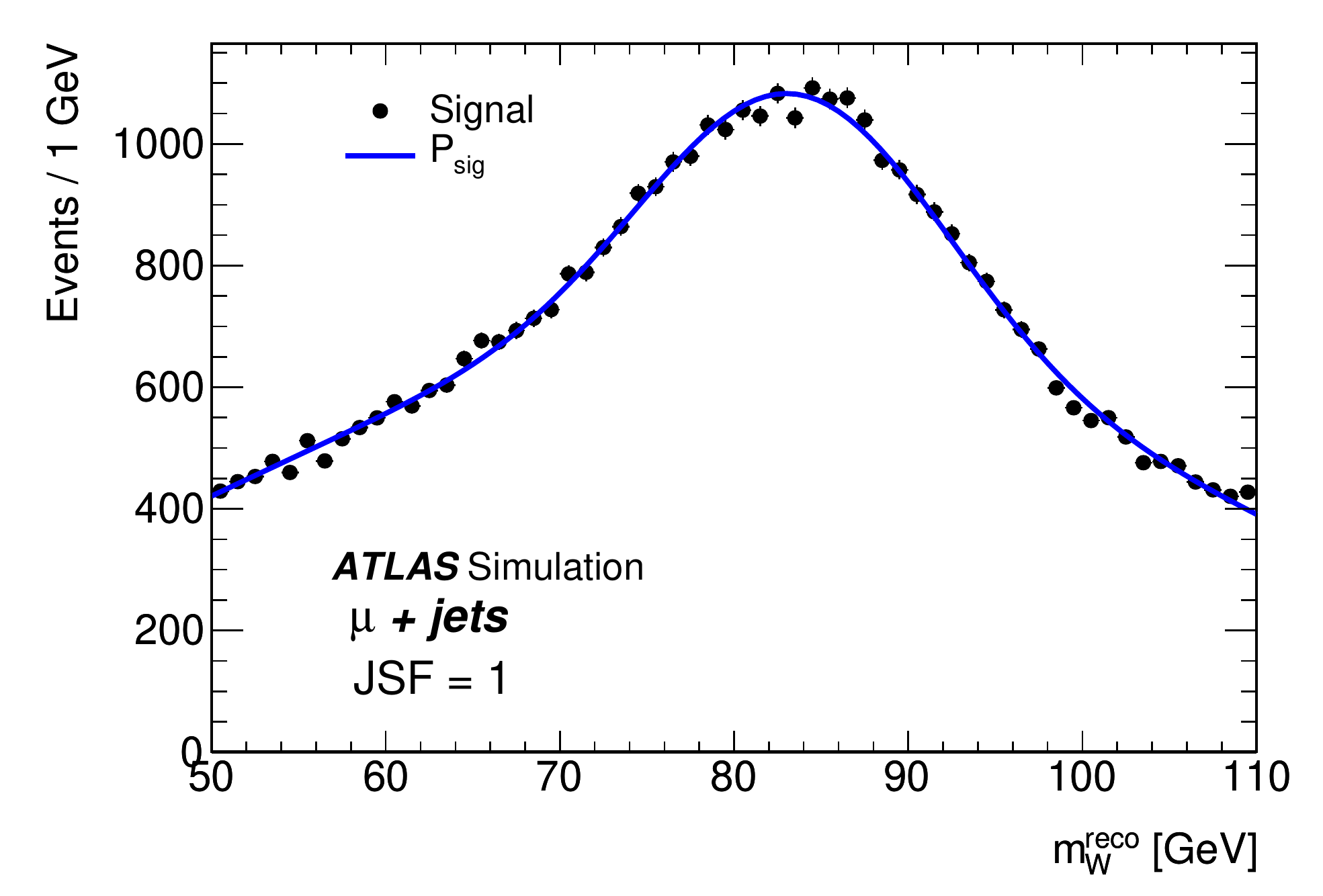}}
\subfigure[\mjets\ channel]{
  \includegraphics[width=0.47\textwidth]{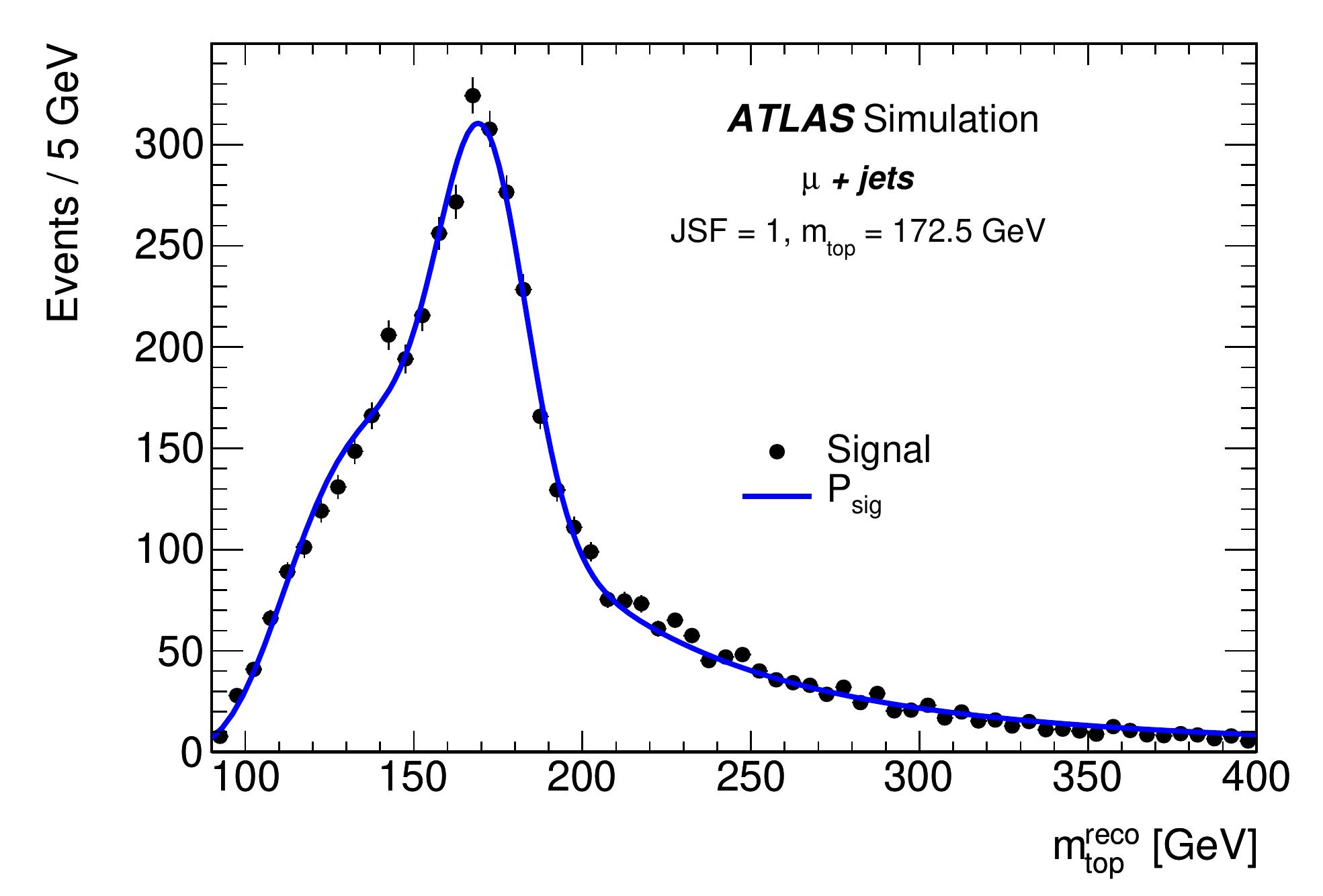}}
\subfigure[\mjets\ channel]{
  \includegraphics[width=0.47\textwidth]{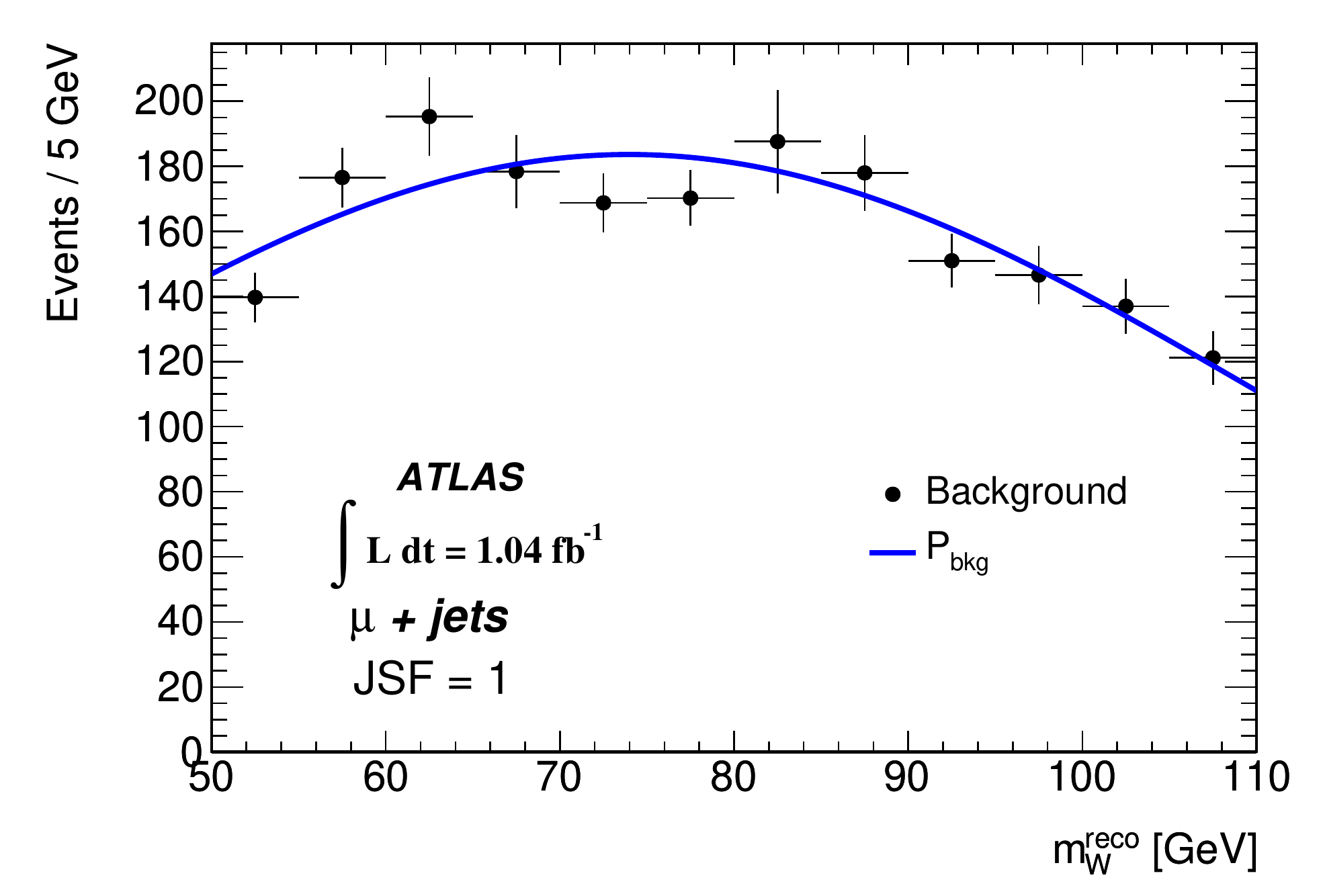}}
\subfigure[\mjets\ channel]{
  \includegraphics[width=0.47\textwidth]{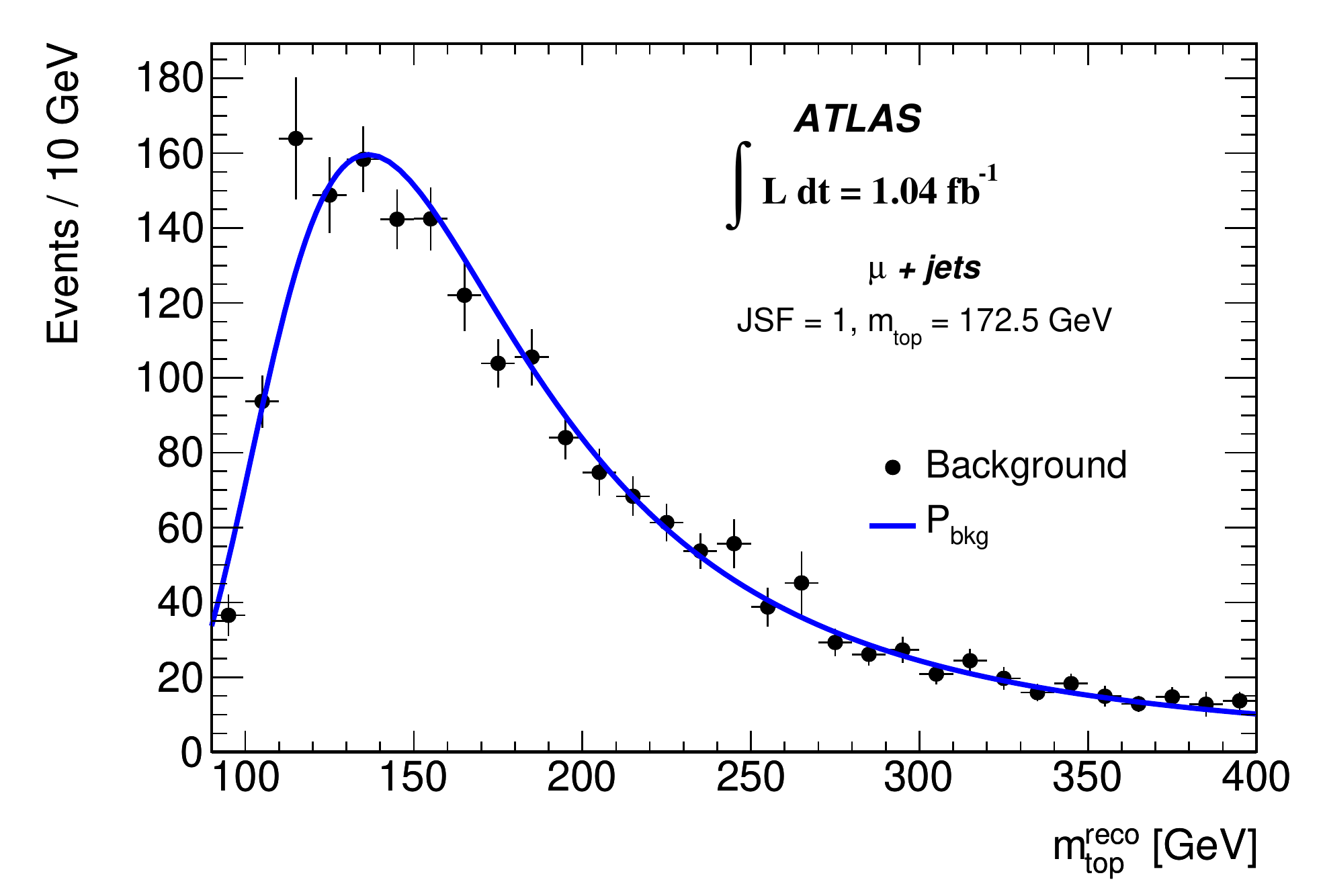}}
\caption{{\bf \td:} Template parameterisations for the \mWr\ and
  \mtr\ distributions for signal and background events for the \mjets\ channel.
  Shown are (a, b) the \mWr\ and \mtr\ distributions for signal events, and (c,
  d) the corresponding distributions for background events, see
  Table~\protect\ref{tab:cutflow}. All distributions are for JSF=1.
  \label{fig:tdtempl}}
\end{figure*}
%
 For the light jet pair, i.e.~for the hadronic \Wboson\ boson decay candidates,
 a kinematic fit is then performed by minimising the following \chiq:
%
\begin{eqnarray*}
\chiq  &=&  \sum_{i=1}^{2}
           \left[\frac{E_\mathrm{jet,i}(1-\ali{i})}{\sigma(E_{\mathrm{jet,i}})}\right]^{2} +
           \left[\frac{M_\mathrm{jet,jet} (\ali{1},\ali{2}) - \mW}{\GW}\right]^2\,,
\end{eqnarray*}
%
 with respect to parton scale factors (\ali{i}) for the jet energies.
 The \chiq\ comprises two components.
 The first component is the sum of squares of the differences of the measured
 and fitted energies of the two reconstructed light jets, $E_\mathrm{jet,i}$,
 individually divided by the squares of their \pt- and \eta-dependent
 resolutions obtained from Monte Carlo simulation, $\sigma(E_{\mathrm{jet,i}})$.
 The second term is the difference of their two-jet invariant mass,
 $M_\mathrm{jet,jet}$, and \mW, divided by the \Wboson\ boson width.
 From these jets the two observables \mWr\ and \mtr\ are constructed.
 The \mWr\ is calculated using the reconstructed light jet four-vectors
 (i.e.~jet energies are not corrected using \ali{i}), retaining the full
 sensitivity of \mWr\ to the JSF.
 In contrast,  \mtr\ is calculated from  these light jet  four-vectors scaled to
 the parton level (i.e.~jet energies  are corrected using \ali{i}) and the above
 determined \bjet.  In this way light jets  in \mtr\ exhibit a  much reduced JES
 sensitivity by construction,  and only the \bjet\ is  directly sensitive to the
 JES.
 The \mWr\ and \mtr\ distributions are shown in Figure~\ref{fig:tdmass} for both
 lepton channels, together with the predictions for signal and
 background. These, in both cases describe the observed distributions well. The
 correlation of these two observables is found to be small for data and
 predictions, and amounts to about \mwmtopcortd.
 \end{sloppypar}

 Templates are constructed for \mtr\ as a function of an input top quark mass in
 the range 160~\GeV~--~190~\GeV, and of an input value for the JSF in the range
 0.9~--~1.1, and, finally, for \mWr\ as a function of the assumed JSF for the
 same range.
 The signal templates for the \mWr\ and \mtr\ distributions, shown for the
 \mjets\ channel and for JSF=1 in Figure~\ref{fig:tdtempl}(a)
 and~\ref{fig:tdtempl}(b), are fitted to a sum of two Gaussian functions for
 \mWr, and to the sum of a Gaussian and a Landau function for \mtr.
 Since, for this analysis, the background templates are constructed including
 single top quark production events, the background fit for the
 \mtr\ distribution is assumed to be \mt\ dependent. For the background, the
 \mWr\ distribution, again shown for the \mjets\ channel in
 Figure~\ref{fig:tdtempl}(c), is fitted to a Gaussian function and the
 \mtr\ distribution, Figure~\ref{fig:tdtempl}(d), to a Landau function.
 For all parameters of the functions that also depend on the JSF, a linear
 parameterisation is chosen.
 The quality of all fits is good for the signal and background contributions and
 for both channels.

 Signal and background probability density functions for the \mtr\ and
 \mWr\ distributions are used in an unbinned likelihood fit to the data for all
 events, $i=1,\dots N$. The likelihood function maximised is:
%
\begin{eqnarray*}
 \Like{shape}(\mWr, \mtr \vert\mt, \JSF, \nbkg)  &=& 
\nonumber \\ 
 \prod_{i=1}^{N} \Ptop(\mtr\vert\mt, \JSF, \nbkg)_{i} 
 &\times& \nonumber \\ 
 \quad\PW(\mWr\vert\JSF, \nbkg)_{i}& & \,\,,
\end{eqnarray*}
\noindent
with:
\begin{eqnarray*}
 \Ptop &=& (N-\nbkg)\cdot\Ptopsig(\mtr\vert\mt, \JSF)_{i} +
 \nonumber \\&&\quad\quad\quad
 \nbkg\cdot\Ptopbkg(\mtr\vert\mt, \JSF)_{i}\,\,,
 \nonumber \\
   \PW &=& (N-\nbkg)\cdot\PWsig(\mWr\vert\JSF)_{i} +
 \nonumber \\
       && \quad\quad\quad \nbkg\cdot\PWbkg(\mWr\vert\JSF)_{i}\,\,.
 \nonumber
\end{eqnarray*}
%
 The three parameters to be determined by the fit are \mt, the \JSF\ and \nbkg.
 Using pseudo-experiments, a good linearity is found between the input top quark
 mass used to perform the pseudo-experiments, and the result of the fits.
 The residual dependence of the reconstructed \mt\ is about 0.1~\GeV\ for a JSF
 shift of 0.01 for both channels, which results in a residual systematic
 uncertainty due to the JES.
 Within their statistical uncertainties, the mean values and widths of the pull
 distributions are consistent with the expectations of zero and one,
 respectively.
 Finally, the expected statistical plus JSF uncertainties (mean $\pm$ RMS)
 obtained from pseudo-experiments at an input top quark mass of
 $\mt=172.5$~\GeV, and for a luminosity of 1~\ifb, are \mtstaeltd~\GeV\ and
 \mtstamutd~\GeV\ for the \ejets\ and \mjets\ channel, respectively.
%
\section{Top quark mass measurement}
\label{sec:measure}
%
\subsection{Evaluation of systematic uncertainties}
\label{sec:syserr}
%
 \begin{sloppypar}
 Each source of uncertainty considered is investigated, when possible, by
 varying the respective quantities by $\pm1\sigma$ with respect to the default
 value. Using the changed parameters, pseudo-experiments are either performed
 directly or templates are constructed and then used to generate
 pseudo-experiments, without altering the probability density function
 parameterisations.
 The difference of the results for \mt\ compared to the standard analysis is
 used to determine the systematic uncertainties.
 For the \td, in any of the evaluations of the systematic uncertainties, apart
 from the JES variations, the maximum deviation of the JSF from its nominal
 fitted value is \JSFmax.
 \end{sloppypar}

 \begin{sloppypar}
 All sources of systematic uncertainties investigated, together with the
 resulting uncertainties, are listed in Table~\ref{tab:results}.
 The statistical precision on \mt\ obtained from the Monte Carlo samples is
 between \mtstasysmin~\GeV\ and \mtstasysmax~\GeV, depending on the available
 Monte Carlo statistics.
 For some sources, pairs of statistically independent samples are used. For
 other sources, the same sample is used, but with a changed parameter. In this
 case the observed \mt\ values for the central and the changed sample are
 statistically highly correlated.
 In all cases, the actual observed difference is quoted as the systematic
 uncertainty on the corresponding source, even if it is smaller than the
 statistical precision of the difference.
 The total uncertainty is calculated as the quadratic sum of all individual
 contributions, i.e.~neglecting possible correlations.
 The estimation of the uncertainties from the individual contributions is
 described in the following.
 \end{sloppypar}

 {\bf Jet energy scale factor:} This is needed to separate the quoted
 statistical uncertainty on the result of the \td\ into a purely statistical
 component on \mt\ analogous to the one obtained in an 1d-analysis, and the
 contribution stemming from the simultaneous determination of the JSF.
 This uncertainty is evaluated for the \td\ by in addition performing a
 one-dimensional (i.e.~JSF-constraint) fit to the data, with the JSF fixed to
 the value obtained in the two-dimensional fit. The quoted statistical precision
 on \mt\ is the one from the one-dimensional fit. The contribution of the JSF is
 obtained by quadratically subtracting the statistical uncertainties on \mt\ for
 the one-dimensional and two-dimensional fit of the \td.

 {\bf Method calibration:} The limited statistics of the Monte Carlo samples
 leads to a systematic uncertainty in the template fits, which is reflected in
 the residual mass differences between the fitted and the input mass for a given
 Monte Carlo sample. The average difference observed in the six samples with
 different input masses is taken as the uncertainty from this source.
%
\begin{table*}[tbp!]
\begin{center}
\begin{tabular}{|l|r|r|r|r||r|r|c|}\cline{2-8}
\multicolumn{1}{c|}{}  & \multicolumn{2}{c|}{\od} & \multicolumn{2}{c||}{\td } & 
\multicolumn{2}{c|}{Combinations} & Correlation \\\cline{2-8}
\multicolumn{1}{c|}{}                 & \ejets & \mjets & \ejets & \mjets & 1d & 2d & \rof \\\hline
Measured value of \mt\                & \mtatoodelval & \mtatoodmuval  & \mtatotdelval & \mtatotdmuval
                                      & \mtodval      & \mttdval       & \\\hline
Data statistics                       & \mtatoodelsta & \mtatoodmusta  & \mtatotdelsta & \mtatotdmusta
                                      & \mtodsta      & \mttdsta       &   \\
Jet energy scale factor               &     na &     na &   0.59 &   0.51 &     na &   0.43 & 0   \\
Method calibration                    &   0.07 &$<0.05$ &   0.10 &$<0.05$ &$<0.05$ &   0.07 & 0   \\\hline
Signal MC generator                   &   0.81 &   0.69 &   0.39 &   0.22 &   0.74 &   0.33 & 1   \\
Hadronisation                         &   0.33 &   0.52 &   0.20 &   0.06 &   0.43 &   0.15 & 1   \\
Pileup                                &$<0.05$ &$<0.05$ &$<0.05$ &$<0.05$ &$<0.05$ &$<0.05$ & 1   \\
Underlying event                      &   0.06 &   0.10 &   0.42 &   0.96 &   0.08 &   0.59 & 1   \\
Colour reconnection                   &   0.47 &   0.74 &   0.32 &   1.04 &   0.62 &   0.55 & 1   \\
ISR and FSR (signal only)             &   1.45 &   1.40 &   1.04 &   0.95 &   1.42 &   1.01 & 1   \\
Proton PDF                            &   0.22 &   0.09 &   0.10 &   0.10 &   0.15 &   0.10 & 1   \\\hline
\Wj\ background normalisation         &   0.16 &   0.19 &   0.34 &   0.44 &   0.18 &   0.37 & 1   \\
\Wj\ background shape                 &   0.11 &   0.18 &   0.07 &   0.22 &   0.15 &   0.12 & 1   \\
QCD multijet background normalisation &   0.07 &$<0.05$ &   0.25 &   0.33 &$<0.05$ &   0.20 & (1) \\
QCD multijet background shape         &   0.14 &   0.12 &   0.38 &   0.30 &   0.09 &   0.27 & (1) \\\hline
Jet energy scale                      &   1.21 &   1.25 &   0.63 &   0.71 &   1.23 &   0.66 & 1   \\
\bjet\ energy scale                   &   1.09 &   1.21 &   1.61 &   1.53 &   1.16 &   1.58 & 1   \\
\btag\ efficiency and mistag rate     &   0.21 &   0.13 &   0.31 &   0.26 &   0.17 &   0.29 & 1   \\
Jet energy resolution                 &   0.34 &   0.38 &   0.07 &   0.07 &   0.36 &   0.07 & 1   \\
Jet reconstruction efficiency         &   0.08 &   0.11 &$<0.05$ &$<0.05$ &   0.10 &$<0.05$ & 1   \\
Missing transverse momentum           &$<0.05$ &$<0.05$ &   0.12 &   0.16 &$<0.05$ &   0.13 & 1   \\\hline
Total systematic uncertainty          & \mtatoodelsys & \mtatoodmusys  & \mtatotdelsys & \mtatotdmusys 
                                      & \mtodsys      & \mttdsys       &                   \\
Total uncertainty                     &   2.86 &   2.80 &   2.46 &   2.68 &  2.66 &   2.39 &    \\

\hline
\end{tabular}
\end{center} 
\caption{The measured values of \mt\ and the contributions of various sources to
  the uncertainty of \mt\ (in \GeV) together with the assumed correlations
  \rof\ between analyses and lepton channels. Here `0' stands for uncorrelated,
  `1' for fully correlated between analyses and lepton channels, and `(1)' for
  fully correlated between analyses, but uncorrelated between lepton
  channels. The abbreviation 'na' stands for not applicable.
 The combined results described in Section~\protect\ref{sec:rescom}
 are also listed.
 \label{tab:results}}
\end{table*}
%

 {\bf Signal Monte Carlo generator:} The systematic uncertainty related to the
 choice of the generator program is accounted for by comparing the results of
 pseudo-experiments performed with either the \Mcatnlo\ or the
 \Powheg\ samples~\cite{FRI-0701} both generated with $\mt=172.5$~\GeV.

 \begin{sloppypar}
 {\bf Hadronisation:} Signal samples for $\mt=172.5$~\GeV\ from the
 \Powheg\ event generator are produced with either the \Pythia~\cite{SJO-0601}
 or \Herwig~\cite{COR-0001} program performing the hadronisation.
 One pseudo-experiment per sample is performed and the full difference of the
 two results is quoted as the systematic uncertainty.
 \end{sloppypar}

 {\bf Pileup:} To investigate the uncertainty due to additional proton-proton
 interactions which may affect the jet energy measurement, on top of the
 component that is already included in the JES uncertainty discussed below, the
 fit is repeated in data and simulation as a function of the number of
 reconstructed vertices.
 Within statistics, the measured \mt\ is independent of the number of
 reconstructed vertices. This is also observed when the data are instead divided
 into data periods according to the average numbers of reconstructed vertices.
 In this case, the subsets have varying contributions from pileup from preceding
 events.

 However, the effect on \mt\ due to any residual small difference between data
 and simulation in the number of reconstructed vertices was assessed by
 computing the weighted sum of a linear interpolation of the fitted masses as a
 function of the number of primary vertices.
 In this sum the weights are the relative frequency of observing a given number
 of vertices in the respective sample.
 The difference of the sums in data and simulation is taken as the uncertainty
 from this source.

 {\bf Underlying event:} This systematic uncertainty is obtained by comparing
 the \Acermc~\cite{KER-0201,KER-0401} central value, defined as the average of
 the highest and the lowest masses measured on the ISR/FSR variation samples
 described below, with a dataset with a modified underlying event.

 {\bf Colour reconnection:} The systematic uncertainty due to colour
 reconnection is determined using \Acermc\ with \Pythia\ with two different
 simulations of the colour reconnection effects as described in
 Refs.~\cite{SKA-1001,ALB-0601,BUC-1001}. In each case, the difference in the
 fitted mass between two assumptions on the size of colour reconnection was
 measured. The maximum difference is taken as the systematic uncertainty due to
 colour reconnection.

 {\bf Initial and final state QCD radiation:} Different amounts of initial and
 final state QCD radiation can alter the jet energies and the jet multiplicity
 of the events with the consequence of introducing distortions into the measured
 \mtr\ and \mWr\ distributions.
 This effect is evaluated by performing pseudo-experiments for which signal
 templates are derived from seven dedicated \Acermc\ signal samples in which
 \Pythia parameters that control the showering are varied in ranges that are
 compatible with those used in the Perugia Hard/Soft tune
 variations~\cite{SKA-1001}. The systematic uncertainty is taken as half the
 maximum difference between any two samples.

 Using different observables, the additional jet activity accompanying the jets
 assigned to the top quark decays has been studied.
 For events in which one (both) \Wboson\ bosons from the top quark decays
 themselves decay into a charged lepton and a neutrino, the reconstructed jet
 multiplicities~\cite{CON-2011-142} (the fraction of events with no additional
 jet above a certain transverse momentum~\cite{ATL-2012-033}) are measured.
 The analysis of the reconstructed jet multiplicities is not sufficiently
 precise to constrain the presently used variations of Monte Carlo parameters.
 In contrast, for the ratio analysis~\cite{ATL-2012-033} the spread of the
 predictions caused by the presently performed ISR variations is significantly
 wider than the uncertainty of the data, indicating that the present ISR
 variations are generous.

 {\bf Proton PDF:} The signal samples are generated using the
 CTEQ~6.6~\cite{NAD-0801} proton parton distribution functions, PDFs. These
 PDFs, obtained from experimental data, have an uncertainty that is reflected in
 22~pairs of additional PDF sets provided by the CTEQ group.
 To evaluate the impact of the PDF uncertainty on the signal templates, the
 events are re-weighted with the corresponding ratio of PDFs, and 22~pairs of
 additional signal templates are constructed. Using these templates one
 pseudo-experiment per pair is performed.
 The uncertainty is calculated as half the quadratic sum of differences of the
 22~pairs as suggested in Ref.~\cite{PUM-0201}.

 {\bf $\mathbf{\Wj}$ background normalisation:} The uncertainty on the
 \Wj\ background determined from data is dominated by the uncertainty on the
 heavy flavour content of these events and amounts to \sysWnorm. The difference
 in \mt\ obtained by varying the normalisation by this amount is taken as the
 systematic uncertainty.

 {\bf $\mathbf{\Wj}$ background shape:} The impact of the variation of the shape
 of the \Wj\ background contribution is studied using a re-weighting
 algorithm~\cite{CON-2011-023} which is based on changes observed on stable
 particle jets when model parameters in the \Alpgen\ Monte Carlo program are
 varied.
 
 {\bf QCD multijet background normalisation:} The estimate for the background
 from QCD multijet events determined from data is varied by \uncqnorm to account
 for the current understanding of this background source~\cite{CON-2011-023} for
 the signal event topology.
 
 {\bf QCD multijet background shape:} The uncertainty due to the QCD background
 shape has been estimated comparing the results from two data driven methods,
 for both channels, see Ref.~\cite{CON-2011-023} for details. For this
 uncertainty pseudo-experiments are performed on QCD background samples with
 varied shapes.
 
 {\bf Jet energy scale:} The jet energy scale is derived using information from
 test-beam data, LHC collision data and simulation.
 Since the energy correction procedure involves a number of steps, the JES
 uncertainty has various components originating from the calibration method, the
 calorimeter response, the detector simulation, and the specific choice of
 parameters in the physics model employed in the Monte Carlo event generator.
 The JES uncertainty varies between \uncjesglobmin\ and \uncjesglobmax\ in the
 central region, depending on jet $p_T$ and $\eta$ as given
 in Ref.~\cite{ATL-2011-085}.
 These values include uncertainties in the flavour composition of the sample and
 mis-measurements from jets close by. Pileup gives an additional uncertainty of
 up to \uncpilecent (\uncpileforw) in the central (forward) region.
 Due to the use of the observable \RtW\ for the \od, and to the simultaneous fit
 of the JSF and \mt\ for the \td, which mitigate the impact of the JES on
 \mt\ differently, the systematic uncertainty on the determined \mt\ resulting
 from the uncertainty of the jet energy scale is less than 1$\%$, i.e.~much
 smaller than the JES uncertainty itself.

 {\bf Relative $\mathbf{\bjet}$ energy scale:} This uncertainty is uncorrelated
 with the jet energy scale uncertainty and accounts for the remaining
 differences between jets originating from light quarks and those from
 \bquarks\ after the global JES has been determined.
 For this, an extra uncertainty ranging from \uncbjesmin\ to \uncbjesmax\ and
 depending on jet \pt\ and $\eta$ is assigned to jets arising from the
 fragmentation of \bquarks, due to differences between light jets and gluon
 jets, and jets containing \bhadrons.
 This uncertainty decreases with \pt, and the average uncertainty for the
 spectrum of jets selected in the analyses is below \uncbrel.

 This additional systematic uncertainty has been obtained from Monte Carlo
 simulation and was also verified using \bjets\ in data.
 The validation of the \bjet\ energy scale uncertainty is based on the
 comparison of the jet transverse momentum as measured in the calorimeter to the
 total transverse momentum of charged particle tracks associated to the
 jet. These transverse momenta are evaluated in the data and in Monte Carlo
 simulated events for inclusive jet samples and for
 \bjet\ samples~\cite{ATL-2011-085}.
 Moreover, the jet calorimeter response uncertainty has been evaluated from the
 single hadron response. Effects stemming from \bquark\ fragmentation,
 hadronisation and underlying soft radiation have been studied using different
 Monte Carlo event generation models~\cite{ATL-2011-085}.

 \begin{sloppypar}
 {\bf $\mathbf{\btag}$ efficiency and mistag rate:} The \btag\ efficiency and
 mistag rates in data and Monte Carlo simulation are not identical. To
 accommodate this, \btag\ scale factors, together with their uncertainties, are
 derived per jet~\cite{CON-2011-089,CON-2011-102}.
 They depend on the jet \pt\ and \eta\ and the underlying quark-flavour. For the
 default result the central values of the scale factors are applied, and the
 systematic uncertainty is assessed by changing their values within their
 uncertainties.
 \end{sloppypar}

 {\bf Jet energy resolution:} To assess the impact of this uncertainty, before
 performing the event selection, the energy of each reconstructed jet in the
 simulation is additionally smeared by a Gaussian function such that the width
 of the resulting Gaussian distribution corresponds to the one including the
 uncertainty on the jet energy resolution.
 The fit is performed using smeared jets and the difference to the default
 \mt\ measurement is assigned as a systematic uncertainty.

 {\bf Jet reconstruction efficiency:} The jet reconstruction efficiency for data
 and the Monte Carlo simulation are found to be in agreement with an accuracy of
 better than \uncjeff~\cite{ATL-2011-085}.
 To account for this, jets are randomly removed from the events using that
 fraction. The event selection and the fit are repeated on the changed sample.

 {\bf Missing transverse momentum:} The \met\ is used in the event selection and
 also in the likelihood for the \od, but is not used in the \mt\ estimator for
 either analysis.
 Consequently, the uncertainty due to any mis-calibration is expected to be
 small. The impact of a possible mis-calibration is assessed by changing the
 measured $\met$ within its uncertainty.

 The resulting sizes of all uncertainties are given in Table~\ref{tab:results}.
 They are also used in the combination of results described below.
 The three most important sources of systematic uncertainty for both analyses
 are the relative \bjet\ to light jet energy scale, the modelling of initial and
 final state QCD radiation, and the light jet energy scale.
 Their impact on the precision on \mt\ are different as expected from the
 difference in the estimators used by the two analyses.
%
\subsection{Results}
\label{sec:rescom}
%
 Figure~\ref{fig:odmass} shows the results of the \od\ when performed on data.
%
\begin{figure}[tbp!]
\centering
\subfigure[\ejets\ channel]{
  \includegraphics[width=0.47\textwidth]{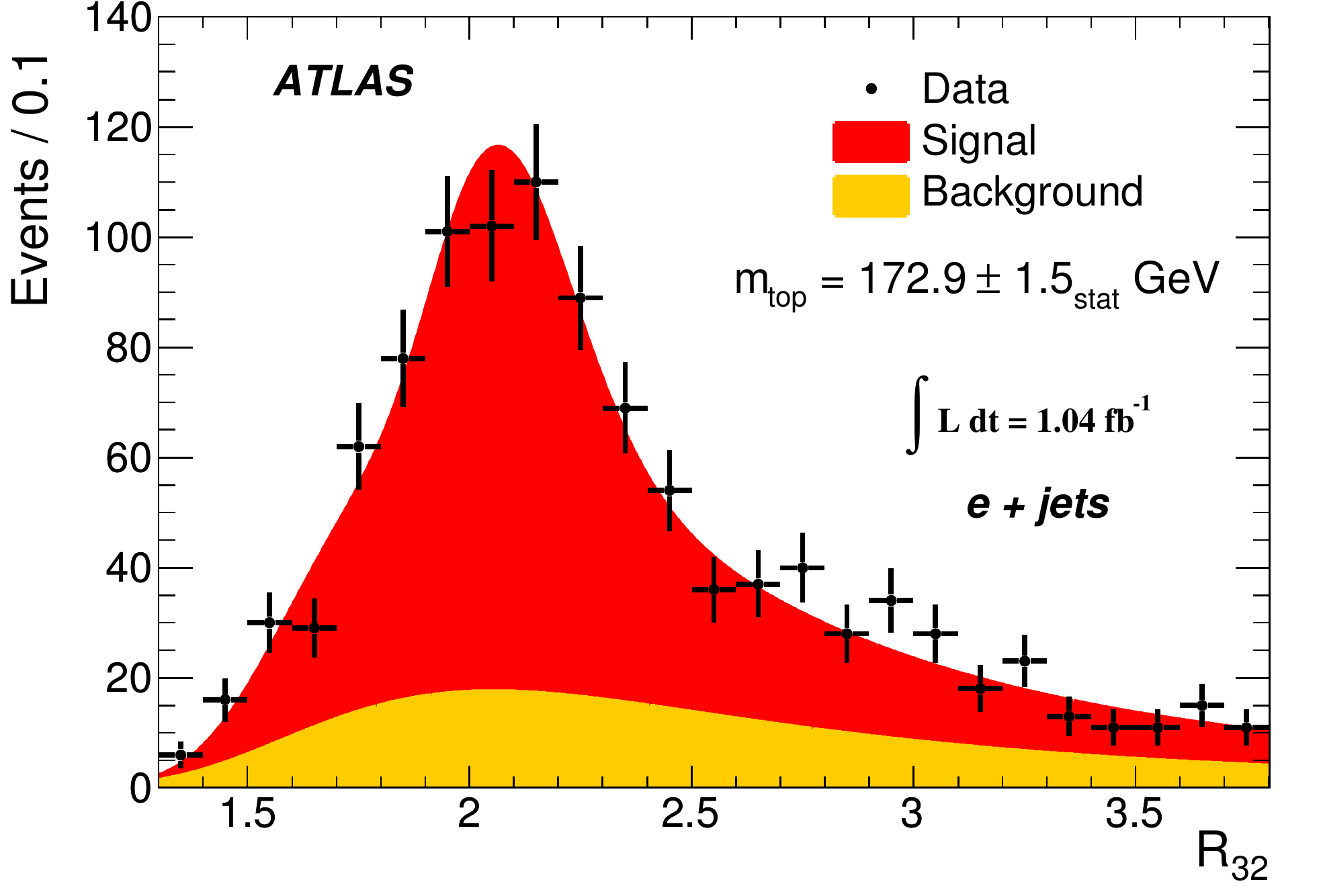}}
\subfigure[\mjets\ channel]{
  \includegraphics[width=0.47\textwidth]{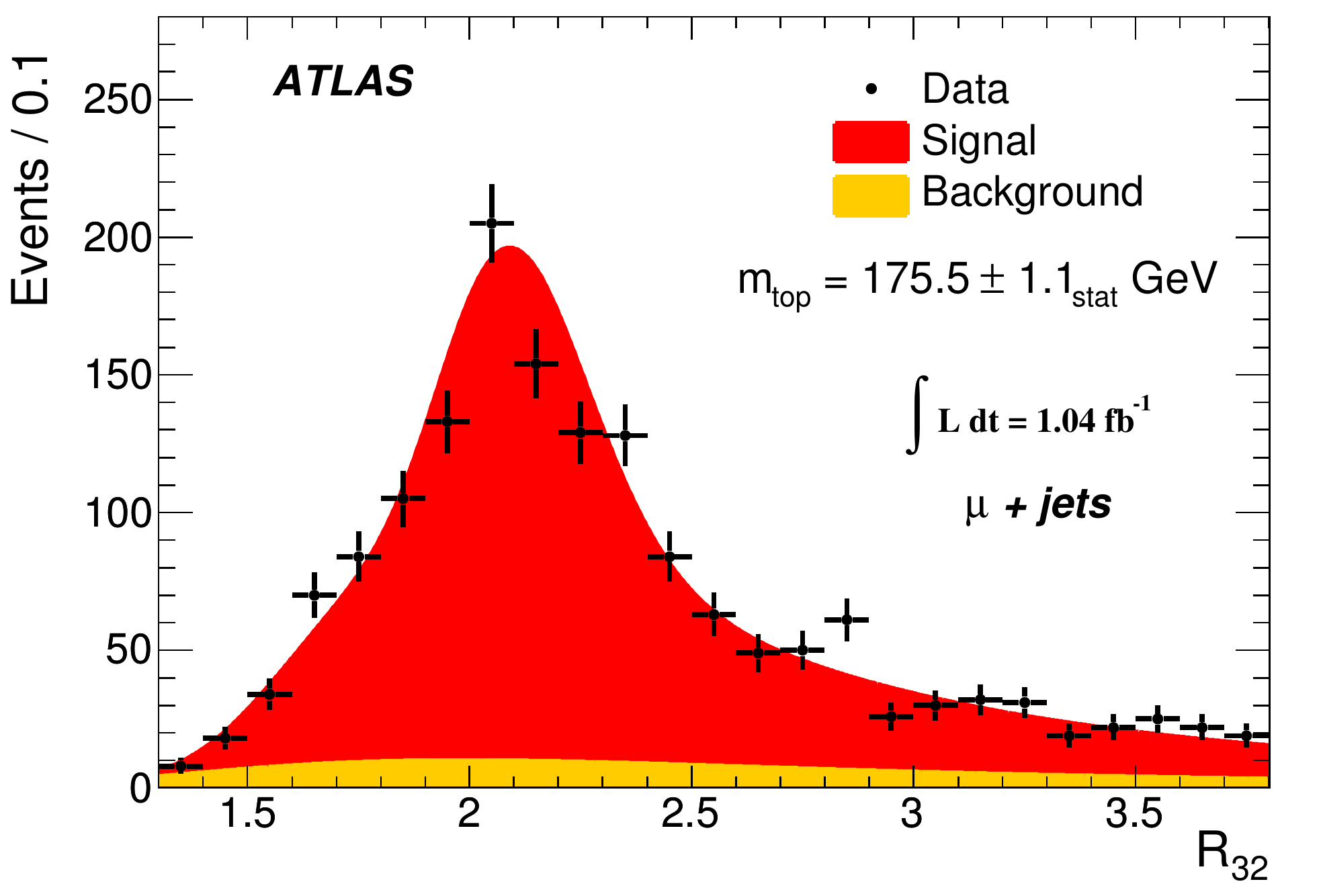}}
\caption{{\bf \od:} The \RtW\ distribution observed in the data together with
  the signal and background contributions determined by the fit. The
  distributions are for (a) the \ejets\ channel and (b) the \mjets\ channel. The
  data points are shown with their statistical uncertainties.
  \label{fig:odmass}}
\end{figure}
%
 For both channels, the fit function describes the data well, with a \chidof\ of
 \chielod/\ndfelod (\chimuod/\ndfmuod) for the \ejets\ (\mjets) channels.
 The observed statistical uncertainties in the data are consistent with the
 expectations given in Section~\ref{sec:1dim} with the \ejets\ channel
 uncertainty being slightly higher than the expected uncertainty of
 \mtstaelod~\GeV.
 The results from both channels are statistically consistent and are:
%
\begin{eqnarray*}
 \mt & = & \mtatoodel\GeV\quad\mbox{(1d \ejets)},\\
 \mt & = & \mtatoodmu\GeV\quad\mbox{(1d \mjets)}.
\end{eqnarray*}

 Figure~\ref{fig:tdmassdata} shows the results of the \td\ when performed on
 data for the \ejets\ and \mjets channels.
 Again the fit functions describe the observed distributions well, with a
 \chidof\ of \chieltd/\ndfeltd (\chimutd/\ndfmutd) for the sum of the \mWr\ and
 \mtr\ distributions in the \ejets\ (\mjets) channels.
 The two-dimensional uncertainty ellipses for both channels are shown in
 Figure~\ref{fig:tdmasscor}.
 The results from both channels are:
%
\begin{eqnarray*}
\centering
 \mt & = & \mtatotdelori\GeV\quad\mbox{(2d \ejets)},\\
 \mt & = & \mtatotdmuori\GeV\quad\mbox{(2d \mjets)}.
\end{eqnarray*}
%
 Within statistical uncertainties these results are consistent with each other,
 and the observed statistical uncertainties in the data are in accord with the
 expectations given in Section~\ref{sec:2dim}, however, for this analysis, with
 the \ejets\ channel uncertainty being slightly lower than the expected
 uncertainty of \mtstaeltd~\GeV.
 The corresponding values for the JSF are $\X{\jesatotdelval}{\jesatotdelsta}$
 and $\X{\jesatotdmdval}{\jesatotdmdsta}$ in the \ejets\ and \mjets\ channels,
 respectively, where the uncertainties are statistical only.
 The JSF values fitted for the two channels are consistent within their
 statistical uncertainty.
 For both channels, the correlation of \mt\ and the JSF in the fits is about
 \cortd.
%
\begin{figure*}[tbp!]
\centering
\subfigure[\ejets\ channel]{
  \includegraphics[width=0.47\textwidth]{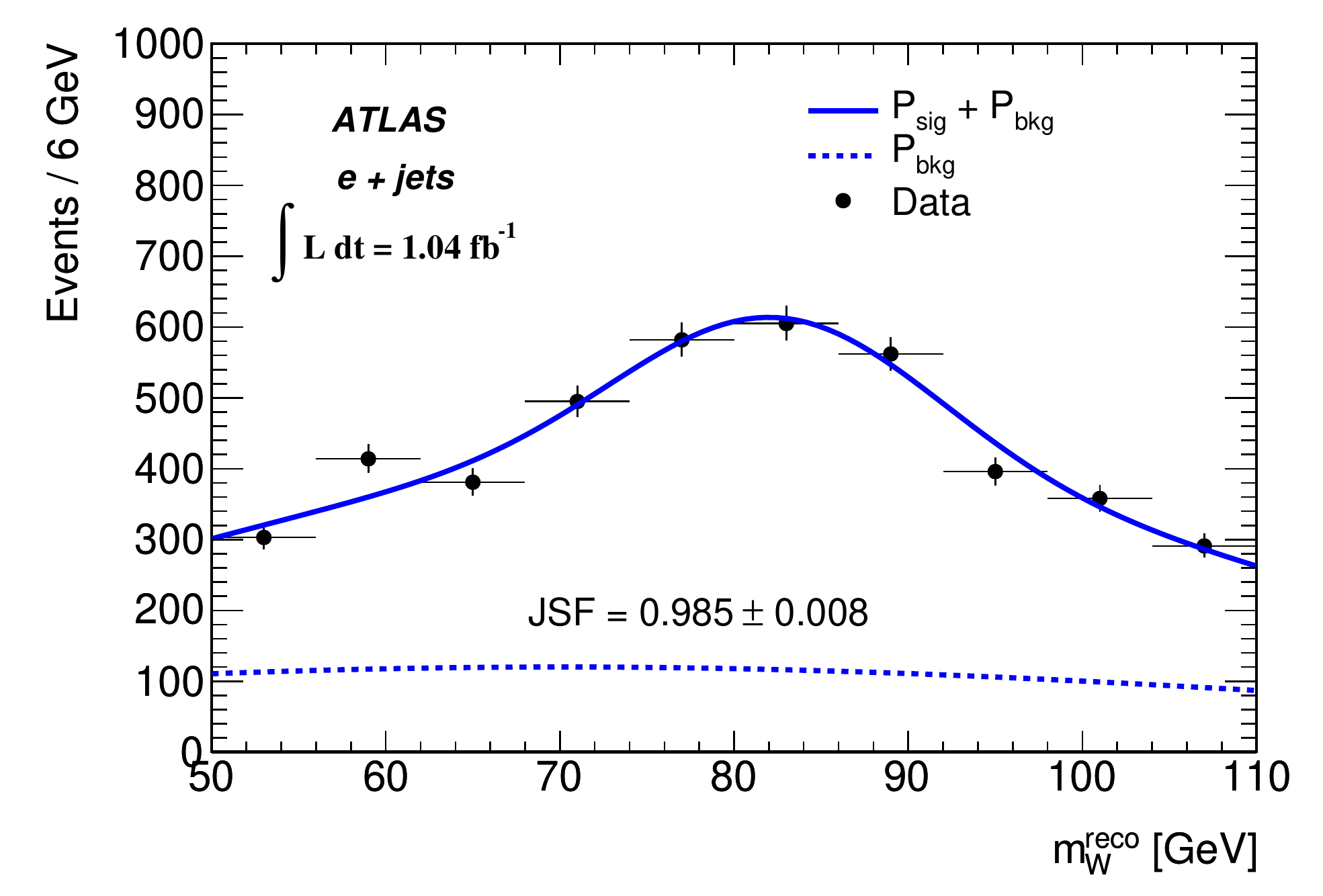}}
\subfigure[\mjets\ channel]{
  \includegraphics[width=0.47\textwidth]{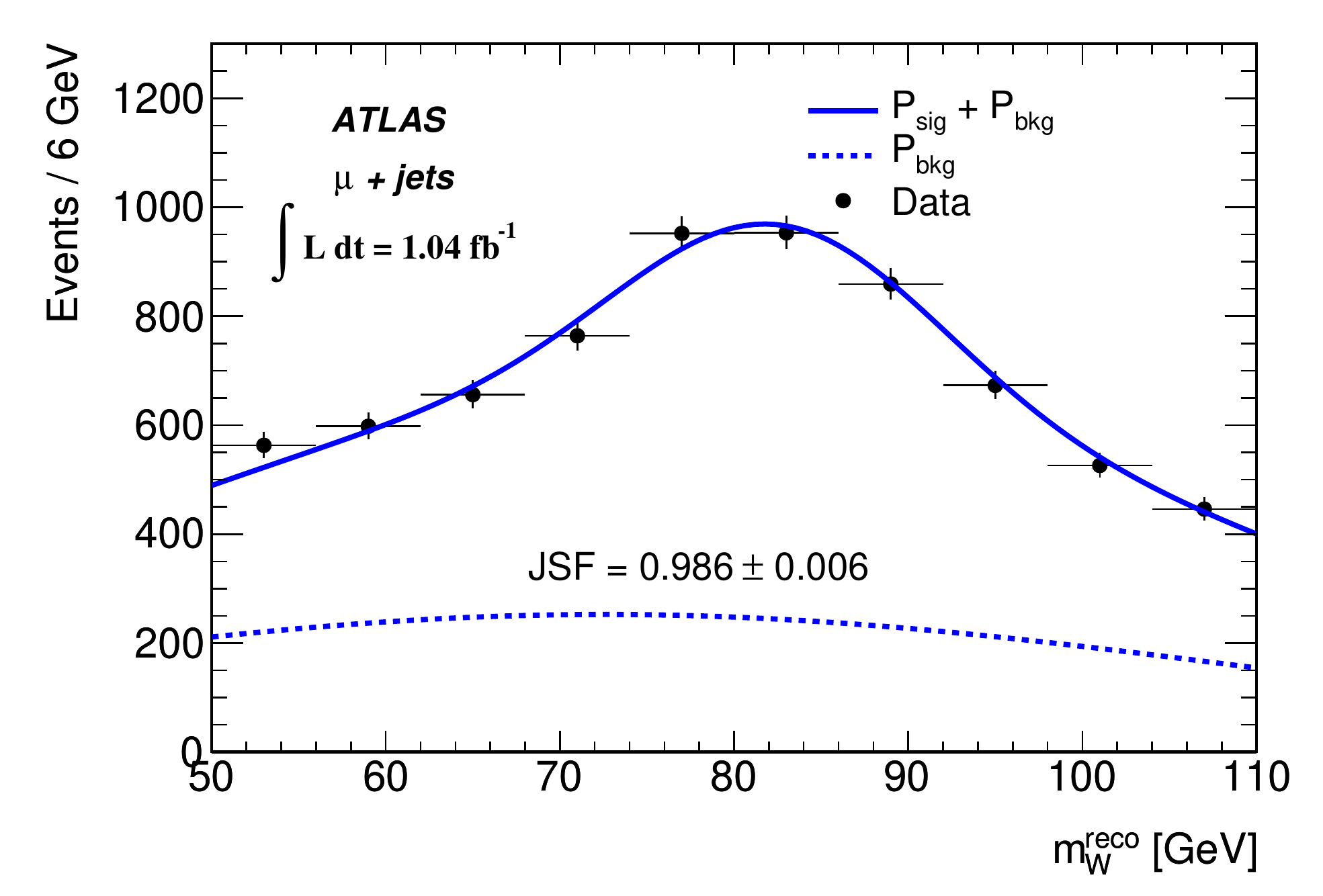}}
\subfigure[\ejets\ channel]{
  \includegraphics[width=0.47\textwidth]{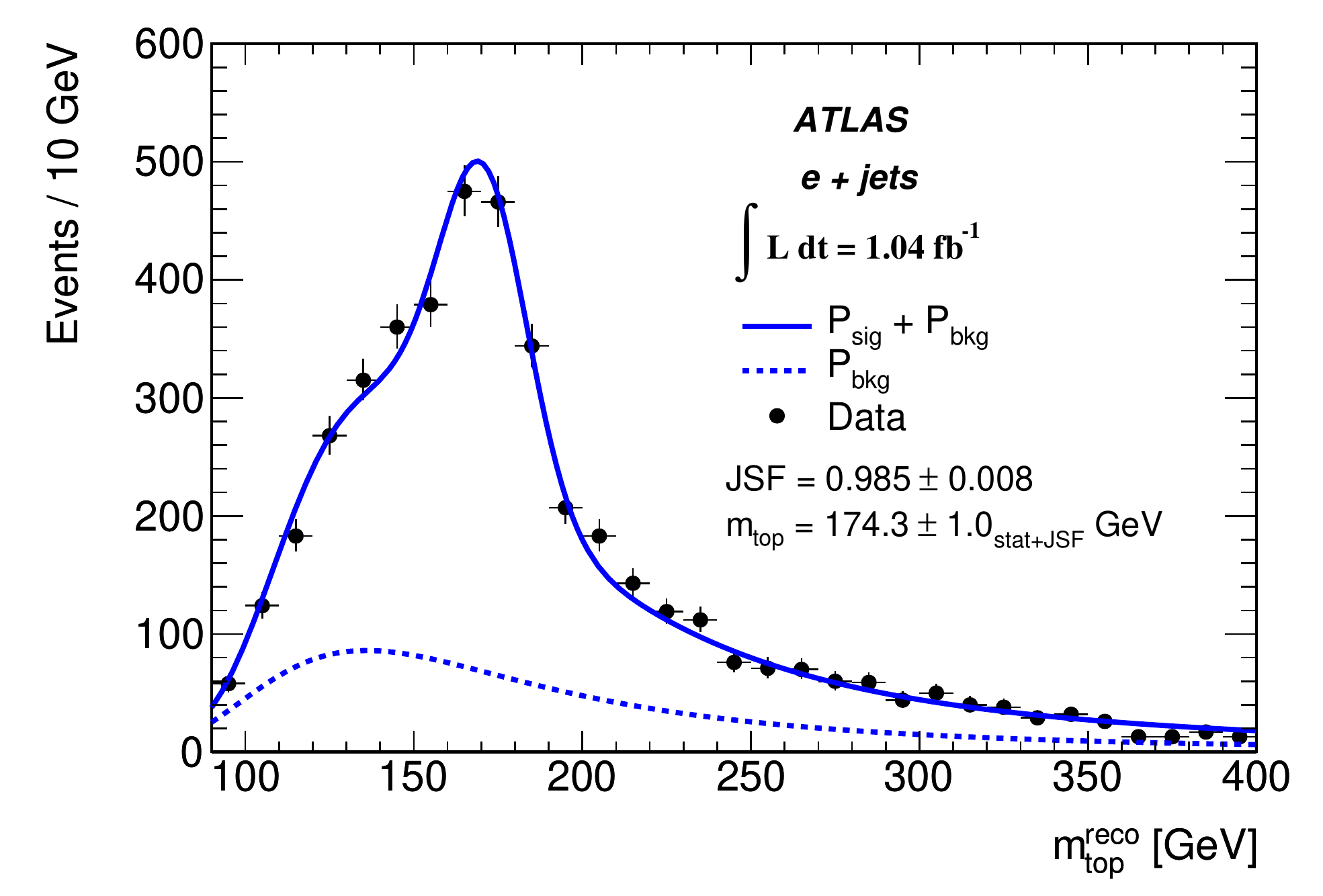}}
\subfigure[\mjets\ channel]{
  \includegraphics[width=0.47\textwidth]{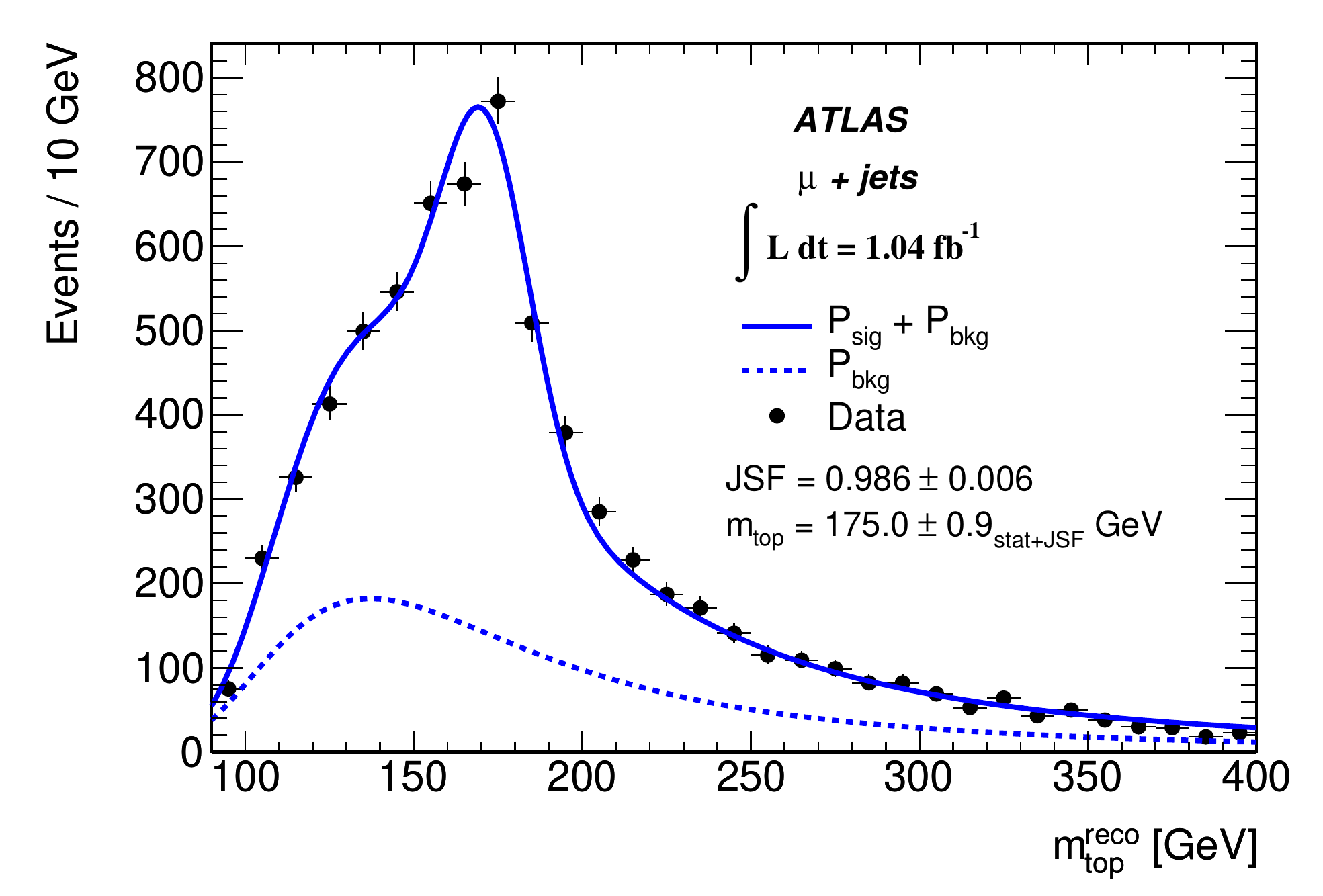}}
\caption{{\bf \td:} Mass distributions fitted to the data for the
  \ejets\ channel on the left and the \mjets\ channel on the right. Shown are
  (a, b) the \mWr\ distributions, and in (c, d) the \mtr\ distributions.  The
  data points are shown with their statistical uncertainties. The lines denote
  the background probability density function (dashed) and the sum of the signal
  and background probability density functions (full).
  \label{fig:tdmassdata}}
\end{figure*}

 When separating the statistical and JSF component of the result as explained in
 the discussion of the JSF uncertainty evaluation in Section~\ref{sec:syserr},
 the result from the \td\ yields:
%
\begin{eqnarray*}
 \mt & = & \mtatotdel\GeV\quad\mbox{(2d \ejets)},\\
 \mt & = & \mtatotdmu\GeV\quad\mbox{(2d \mjets)}.
\end{eqnarray*}
%
 These values together with the breakdown of uncertainties are shown in
 Table~\ref{tab:results} and are used in the combinations.

 \begin{sloppypar}
 Due to the additional event selection requirements used in the \od\ to optimise
 the expected uncertainty described in Section~\ref{sec:addreq}, for both
 channels the \td\ has the smaller statistical uncertainty, despite the better
 top quark mass resolution of the \od.
 Both analyses are limited by the systematic uncertainties, which have different
 relative contributions per source but are comparable in total size, i.e.~the
 difference in total uncertainty between the most precise and the least precise
 of the four measurements is only \difftot.
 \end{sloppypar}

 The four individual results are all based on data from the first part of the
 2011 data taking period.
 The \ejets\ and \mjets\ channel analyses exploit exclusive event selections and
 consequently are statistically uncorrelated within a given analysis.
 In contrast, for each lepton channel the data samples partly overlap, see
 Section~\ref{sec:evsel}.
 However, because the selection of the jet triplet and the construction of the
 estimator of \mt\ are different, the two analyses are less correlated than the
 about \fraco\ that would be expected from the overlap of events.

 The statistical correlation of the two results for each of the lepton channels
 is evaluated using the Monte Carlo method suggested in Ref.~\cite{LYO-1988},
 exploiting the large Monte Carlo signal samples.
 For all four measurements (two channels and two analyses), five hundred
 independent pseudo-experiments are performed, ensuring that for every single
 pseudo-experiment the identical events are input to all measurements.
 The precision of the determined statistical correlations depends purely on the
 number of pseudo-experiments performed, and in particular, it is independent of
 the uncertainty of the measured \mt\ per pseudo-experiment.
 In this analysis, the precision amounts to approximately 4$\%$ absolute,
 i.e.~this estimate is sufficiently precise that its impact on the uncertainty
 on \mt, given the low sensitivity of the combined results of \mt\ to the
 statistical correlation, is negligible.
 For the \od, the signal is comprised of \ttbar\ and single top quark
 production, whereas for the \td\ the single top quark production process is
 included in the background, see Table~\ref{tab:cutflow}.
 Consequently, the \Mcatnlo\ samples generated at $\mt=172.5$~\gev\ for both
 processes are used appropriately for each analysis in determining the
 statistical correlations.
 The statistical correlation between the results of the two analyses is
 \statcorel\ (\statcormu) in the \ejets (\mjets) channels, respectively.
 Given these correlations, the two measurements for each lepton channel are
 statistically consistent for both lepton flavours.
%
\begin{figure}[tbp!]
\centering
\subfigure[\ejets\ channel]{
  \includegraphics[width=0.47\textwidth]{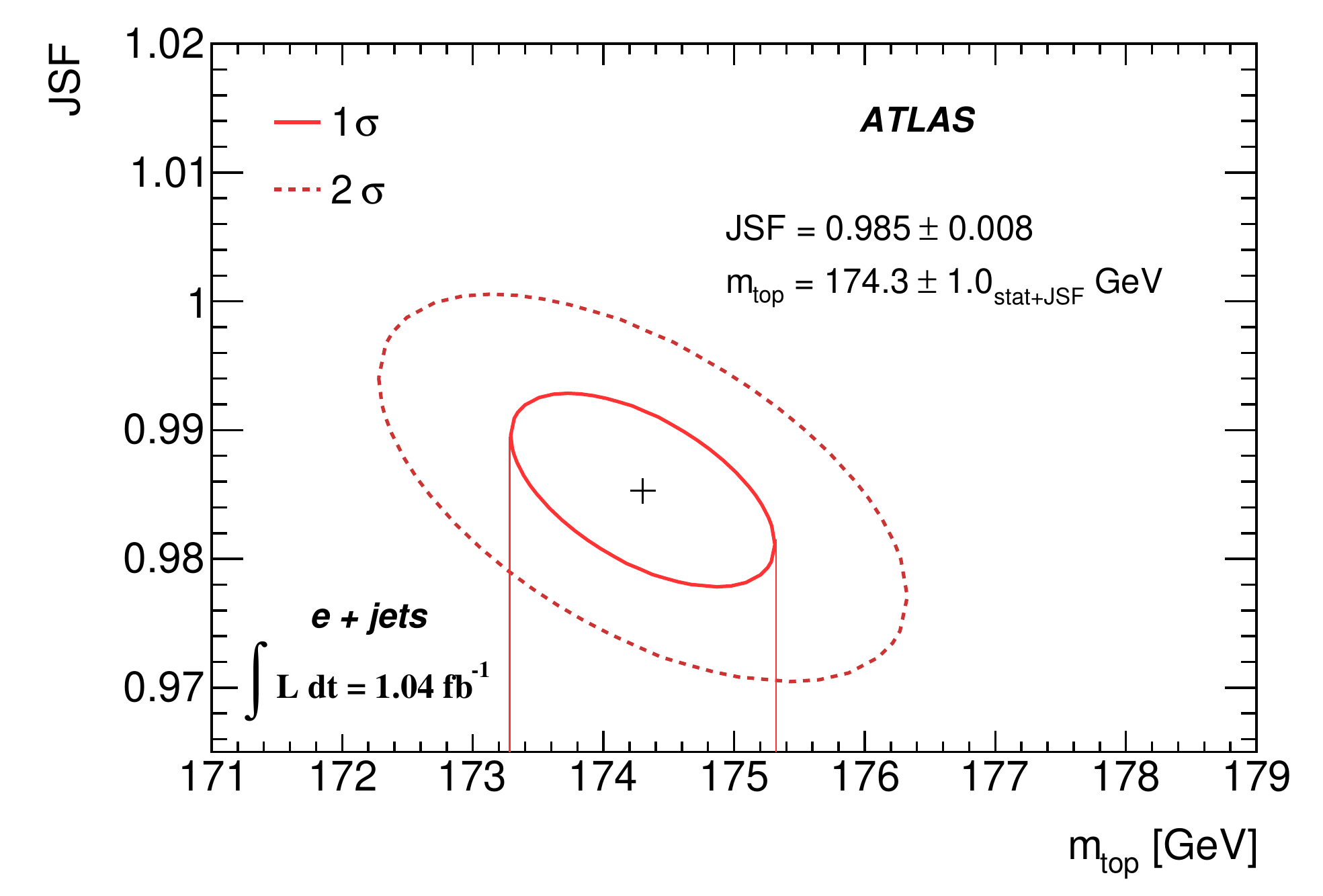}}
\subfigure[\mjets\ channel]{
  \includegraphics[width=0.47\textwidth]{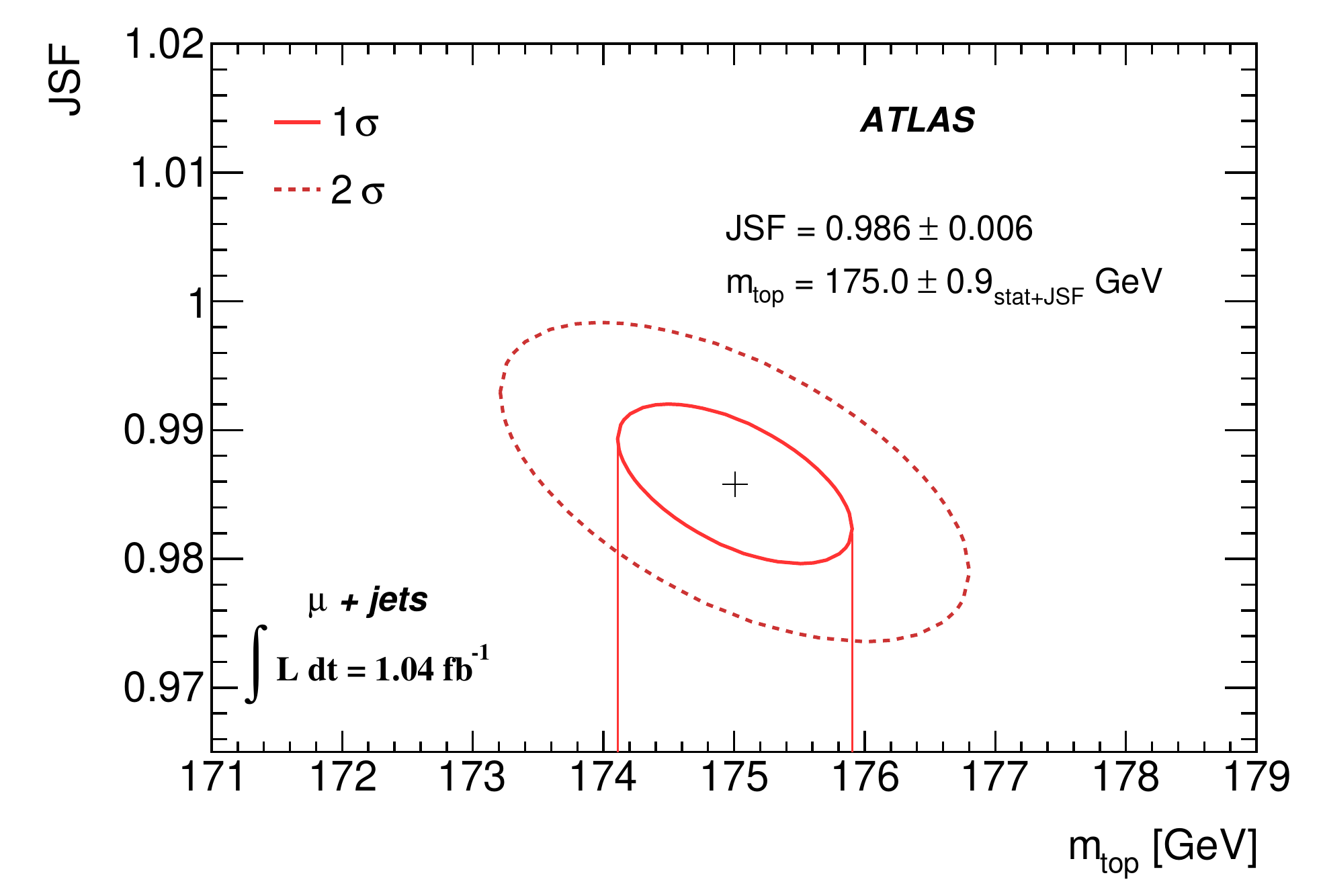}}
\caption{{\bf \td:} The correlation of the measured top quark mass \mt, and jet
  energy scale factor JSF for (a) the \ejets\ channel, and (b) the
  \mjets\ channel. The ellipses correspond to the one- and two standard
  deviation uncertainties of the two parameters.
  \label{fig:tdmasscor}}
\end{figure}
%

 The combinations of results are performed for the individual measurements and
 their uncertainties listed in Table~\ref{tab:results} and using the formalism
 described in Refs.~\cite{LYO-1988,VAL-0301}.
 The statistical correlations described above are used.
 The correlations of systematic uncertainties assumed in the combinations fall
 into three classes. For the uncertainty in question the measurements are either
 considered uncorrelated $\rof = 0$, fully correlated between analyses and
 lepton channels $\rof = 1$, or fully correlated between analyses, but
 uncorrelated between lepton channels denoted with $\rof = (1)$.
 A correlation of $\rof = 0$ is used for the sources method calibration and jet
 energy scale factor, which are of purely statistical nature.
 The sources with $\rof = 1$ are listed in Table~\ref{tab:results}.
 Finally, the sources with $\rof = (1)$ are QCD background normalisation and
 shape that are based on independent lepton fake rates  in each lepton channel.

 Combining the results for the two lepton channels separately for each analysis
 gives the following results (note that these two analyses are correlated as
 described above):
%
\begin{eqnarray*}
 \mt & = & \mtod\GeV\quad\mbox{(\od)},\\
 \mt & = & \mttd\GeV\quad\mbox{(\td)}.
\end{eqnarray*}
%
 For the \od\ the \mjets\ channel is more precise, and consequently carries a
 larger weight in the combination, whereas for the \td\ this is
 reversed. However, for both analyses, the improvement on the more precise
 estimate by the combination is moderate, i.e.~a few percent, see
 Table~\ref{tab:results}.

 \begin{sloppypar}
 The pairwise correlation of the four individual results range from \rhomin\ to
 \rhomax, with the smallest correlation between the results from the different
 lepton channels of the different analyses, and the largest correlation between
 the ones from the two lepton channels within an individual analysis.
 The combination of all four measurements of \mt\ yields statistical and
 systematic uncertainties on the top quark mass of \mtatostacut~\GeV\ and
 \mtatosyscut~\GeV, respectively.
 Presently this combination does not improve the precision of the measured top
 quark mass from the \td, which has the better expected total uncertainty.
 Therefore, the result from the \td\ is presented as the final result.
 The two analyses will differently profit from progress on the individual
 systematic uncertainties, which can be fully exploited by the method to
 estimate the statistical correlation of different estimators of \mt\ obtained
 in the same data sample together with the outlined combination procedure.
%
\begin{figure}[tbp!]
\centering
\includegraphics[width=0.47\textwidth]{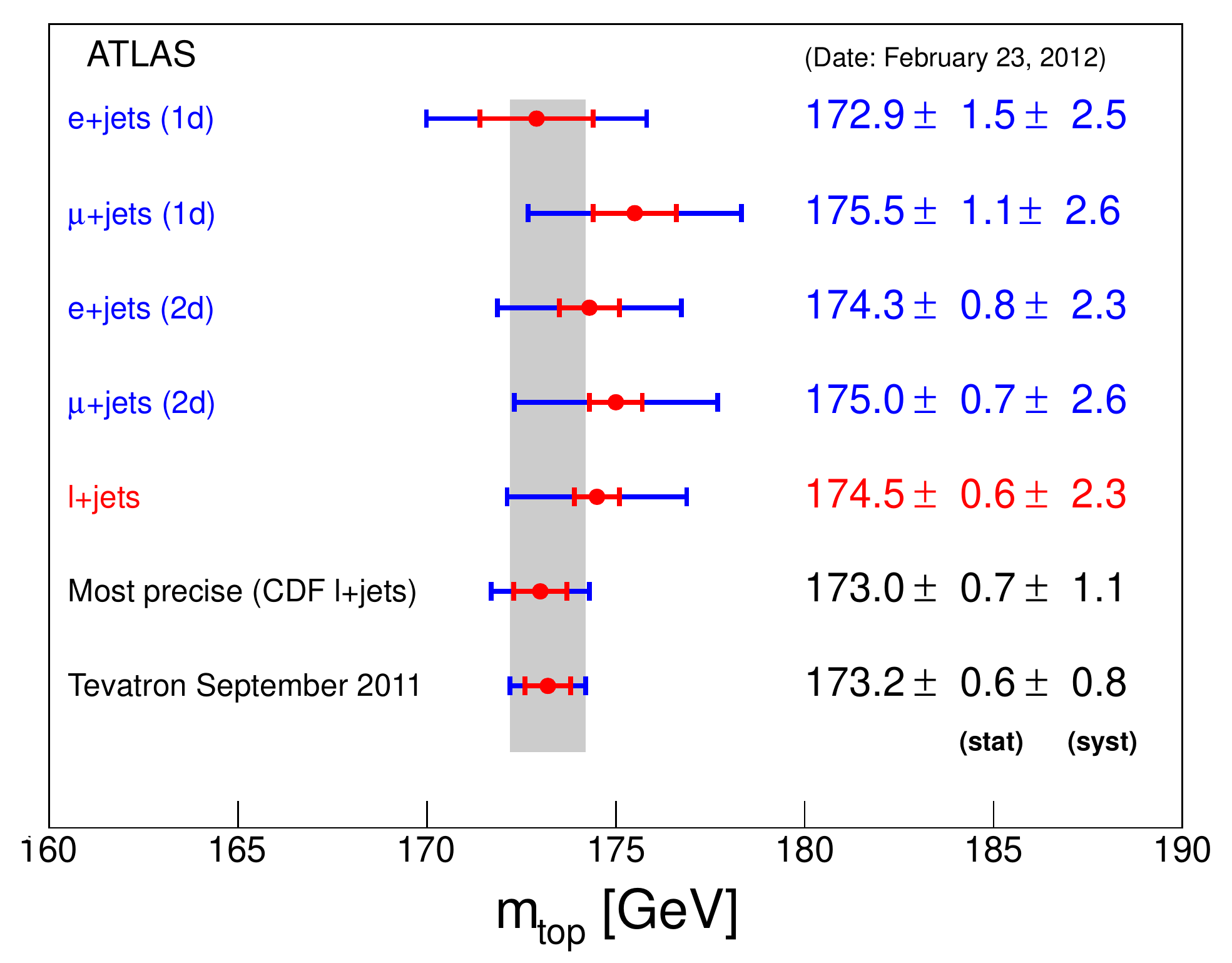}
\caption{The measurements on \mt\ from the individual analyses and the combined
  result from the \td\ compared to the present combined value from the Tevatron
  experiments~\protect\cite{TEV-1101} and to the most precise measurement of
  \mt\ used in that combination.
\label{fig:atlas2011}}
\end{figure}
%
 The results are summarised in Figure~\ref{fig:atlas2011} and compared to
 selected measurements from the Tevatron experiments.
 \end{sloppypar}
%
\section{Summary and conclusion}
\label{sec:summar}
%
 The top quark mass has been measured directly via two implementations of the
 template method in the \ejets\ and \mjets\ decay channels, based on
 proton-proton collision data from 2011 corresponding to an integrated
 luminosity of about \atlumo~\ifb.
 The two analyses mitigate the impact of the three largest systematic
 uncertainties on the measured \mt\ with different methods.
 The \ejets\ and \mjets\ channels, and both analyses, lead to consistent results
 within their correlated uncertainties.

 A combined \od\ and \td\ result does not currently improve the precision of the
 measured top quark mass from the \td\ and hence the \td\ result is presented as
 the final result:
%
\begin{eqnarray*}
\mt&=&\mttd\GeV\,.\nonumber
\end{eqnarray*}
%
 This result is statistically as precise as the \mt\ measurement obtained in the
 Tevatron combination, but the total uncertainty, dominated by systematic
 effects, is still significantly larger.
 In this result, the three most important sources of systematic uncertainty are
 from the relative \bjet\ to light jet energy scale, the modelling of initial
 and final state QCD radiation, and the light quark jet energy scale.
 These sources account for about \maxsys\ of the total systematic uncertainty.
%
%
\section*{Acknowledgements}

We thank CERN for the very successful operation of the LHC, as well as the
support staff from our institutions without whom ATLAS could not be
operated efficiently.

We acknowledge the support of ANPCyT, Argentina; YerPhI, Armenia; ARC,
Australia; BMWF, Austria; ANAS, Azerbaijan; SSTC, Belarus; CNPq and FAPESP,
Brazil; NSERC, NRC and CFI, Canada; CERN; CONICYT, Chile; CAS, MOST and NSFC,
China; COLCIENCIAS, Colombia; MSMT CR, MPO CR and VSC CR, Czech Republic;
DNRF, DNSRC and Lundbeck Foundation, Denmark; ARTEMIS and ERC, European Union;
IN2P3-CNRS, CEA-DSM/IRFU, France; GNAS, Georgia; BMBF, DFG, HGF, MPG and AvH
Foundation, Germany; GSRT, Greece; ISF, MINERVA, GIF, DIP and Benoziyo Center,
Israel; INFN, Italy; MEXT and JSPS, Japan; CNRST, Morocco; FOM and NWO,
Netherlands; RCN, Norway; MNiSW, Poland; GRICES and FCT, Portugal; MERYS
(MECTS), Romania; MES of Russia and ROSATOM, Russian Federation; JINR; MSTD,
Serbia; MSSR, Slovakia; ARRS and MVZT, Slovenia; DST/NRF, South Africa;
MICINN, Spain; SRC and Wallenberg Foundation, Sweden; SER, SNSF and Cantons of
Bern and Geneva, Switzerland; NSC, Taiwan; TAEK, Turkey; STFC, the Royal
Society and Leverhulme Trust, United Kingdom; DOE and NSF, United States of
America.

The crucial computing support from all WLCG partners is acknowledged
gratefully, in particular from CERN and the ATLAS Tier-1 facilities at
TRIUMF (Canada), NDGF (Denmark, Norway, Sweden), CC-IN2P3 (France),
KIT/GridKA (Germany), INFN-CNAF (Italy), NL-T1 (Netherlands), PIC (Spain),
ASGC (Taiwan), RAL (UK) and BNL (USA) and in the Tier-2 facilities
worldwide.
%
%
\bibliographystyle{atlasnote}
\bibliography{main}
\clearpage
\onecolumn
\vspace*{0.2cm}
\input{atlas_authlist}
\end{document}

%% file: atlas_authlist.tex
\begin{flushleft}
{\Large The ATLAS Collaboration}

\bigskip

G.~Aad$^{\rm 48}$,
B.~Abbott$^{\rm 110}$,
J.~Abdallah$^{\rm 11}$,
A.A.~Abdelalim$^{\rm 49}$,
A.~Abdesselam$^{\rm 117}$,
O.~Abdinov$^{\rm 10}$,
B.~Abi$^{\rm 111}$,
M.~Abolins$^{\rm 87}$,
O.S.~AbouZeid$^{\rm 157}$,
H.~Abramowicz$^{\rm 152}$
% ...
\bigskip

% ...
$^{10}$ Institute of Physics, Azerbaijan Academy of Sciences, Baku, Azerbaijan\\
$^{11}$ Institut de F\'isica d'Altes Energies and Departament de F\'isica de la Universitat Aut\`onoma  de Barcelona and ICREA, Barcelona, Spain\\
% ...
$^{48}$ Fakult\"{a}t f\"{u}r Mathematik und Physik, Albert-Ludwigs-Universit\"{a}t, Freiburg i.Br., Germany\\
$^{49}$ Section de Physique, Universit\'e de Gen\`eve, Geneva, Switzerland\\
% ...
$^{87}$ Department of Physics and Astronomy, Michigan State University, East Lansing MI, United States of America\\
% ...
$^{110}$ Homer L. Dodge Department of Physics and Astronomy, University of Oklahoma, Norman OK, United States of America\\
$^{111}$ Department of Physics, Oklahoma State University, Stillwater OK, United States of America\\
% ...
$^{117}$ Department of Physics, Oxford University, Oxford, United Kingdom\\
% ...
$^{152}$ Raymond and Beverly Sackler School of Physics and Astronomy, Tel Aviv University, Tel Aviv, Israel\\
% ...
$^{157}$ Department of Physics, University of Toronto, Toronto ON, Canada\\
% ...
\end{flushleft}
